\documentclass{aa}  
\usepackage{color}
\usepackage[colorinlistoftodos]{todonotes}

\usepackage{graphicx}

\newcommand{\HI}{{\ion{H}{1}}}

\newcommand{\kms}{$\,$km$\,$s$^{-1}$}

\newcommand{\ergs}{$\,$erg$\,$s$^{-1}$}

\newcommand{\mJybeam}{mJy beam$^{-1}$}
\newcommand{\muJybeam}{$\mu$Jy beam$^{-1}$}
\newcommand{\msun}{{$M_\odot$}}
\newcommand{\msunyr}{{$M_\odot$ yr$^{-1}$}}

\newcommand{\lsun}{{$L_\odot$}}

\newcommand{\pks}{{PKS\,1549$-$79}}
\newcommand{\coOne}{{CO(1-0)}}
\newcommand{\coThree}{{CO(3-2)}}
\newcommand{\rde}{$R_{\rm 31}$}
\def\HI{\ion{H}{i}}

\def\OIII{[O{\,\small III}]}

\def\emph#1{{\sl #1}}
\newcommand{\ltsima} {$\; \buildrel < \over \sim \;$}
\newcommand{\gtsima} {$\; \buildrel > \over \sim \;$}
\newcommand{\lta} {\lower.5ex\hbox{\ltsima}}
\newcommand{\gta} {\lower.5ex\hbox{\gtsima}}

\usepackage{txfonts}
\begin{document}

   \title{ALMA observations of PKS~1549--79: \\ a case of feeding and feedback in a young radio quasar 
}

\authorrunning{Oosterloo et al.}
\titlerunning{Feeding and feedback in PKS~1549--79}
\author{Tom Oosterloo\inst{1,2}, Raffaella Morganti\inst{1,2}, Clive Tadhunter \inst{3}, J. B. Raymond Oonk\inst{1,4,5}, Hayley E. Bignall\inst{6}, \\ Tasso Tzioumis\inst{7}, Cormac Reynolds\inst{6}}
\institute{ASTRON, the Netherlands Institute for Radio Astronomy, Oude Hoogeveensedijk 4, 7991 PD, Dwingeloo, The Netherlands. 
\and
Kapteyn Astronomical Institute, University of Groningen, Postbus 800,
9700 AV Groningen, The Netherlands
\and
Department of Physics and Astronomy, University of Sheffield, Sheffield, S7 3RH, UK
\and
Leiden Observatory, Leiden University, Postbus 9513, 2300 RA Leiden
\and
SURFsara, Postbus 94613, 1090 GP Amsterdam, The Netherlands
\and{CSIRO Astronomy and Space Science, Kensington 6151, Australia}
\and {Australia Telescope National Facility, CSIRO, PO Box 76, Epping, NSW 1710, Australia}
}

   \date{Received July 5, 2019; accepted October 7, 2019}
 
  \abstract
   {We present  \coOne\ and \coThree\  Atacama Large Millimeter/submillimeter Array observations of the molecular gas in \pks,  as well as  mm and very long baseline interferometry 2.3-GHz continuum observations of its radio jet. \pks\ is one of the closest young, radio-loud quasars caught in an on-going merger in which the active galactic nucleus (AGN) is in the first phases of its evolution. 
   We detect three structures tracing  the accretion and the outflow of molecular gas: kpc-scale tails of gas accreting onto \pks\ from a merger,  a circumnuclear disc in the inner few hundred parsec, and a very broad ($>$2300 \kms) component detected in \coOne\ at the position of the AGN. 
   Thus, in \pks\ we see the co-existence of accretion and the ejection of gas. 
   The line ratio \coThree/\coOne\ suggests that  the  gas in the circumnuclear-disc  has both high densities and high kinetic temperatures. We estimate a mass outflow rate of at least 650 \msunyr. This massive outflow is  confined to the inner region ($ r < 120$ pc) of the galaxy, which suggests that the AGN drives the outflow.  Considering the amount of molecular gas available in the central nuclear disc and the observed outflow rate, we estimate a time scale of $\sim$10$^5$ yr over which the AGN would be able to destroy the circumnuclear disc, although gas from the merger may come in from larger radii, rebuilding this disc at the same time. 
   The AGN appears  to  self-regulate gas accretion to the centre and onto the super-massive black hole. Surprisingly, from a comparison with Hubble Space Telescope data, we find that the ionised gas outflow is more extended. Nevertheless, the warm outflow is about two orders of magnitude less massive  than the molecular outflow. \pks\ does not seem to follow the scaling relation between bolometric luminosity and the relative importance of warm ionised and molecular outflows  claimed to exist for other AGN. We argue that, although \pks\ hosts a powerful quasar nucleus and an ultra-fast outflow, the radio jet  plays a significant role in producing the outflow, which creates a cocoon of disturbed gas that expands into the circumnuclear disc. 
   }
   \keywords{galaxies: active - galaxies: individual: PKS~1549--79 - ISM: jets and outflow - radio lines: galaxies}
   \maketitle  
%

\section{Introduction}
\label{sec:introduction}

Research in the past twenty years has told us that active super-massive black holes (SMBH) play an important role in galaxy evolution due to the enormous amount of energy they can release. Interestingly, the attempts to understand the impact of such active galactic nuclei (AGN) feedback have brought new, unexpected insights into the physical conditions in the central regions of their host galaxy. The  presence of fast,  massive outflows of atomic  (\HI) and molecular gas   represents an important manifestation of  feedback. 
However, studies of AGN feedback have also shown how complex this process is. This partly reflects the complexity of the AGN phenomenon itself, as well as the variety of the different processes involved in the surrounding gas. Indeed, among the many open-ended questions is the relative role of different types of AGN  (Wylezalek \& Morganti 2018, Harrison et al.\ 2018) and how the energy they release actually couples to the surrounding medium.  Gas outflows are believed to play an important role in this. However, only in a few cases do the outflows seem to go beyond a few kpc distance from the nucleus. This puts into question the impact they may have on regulating  star formation on the scale of the entire galaxy. A better understanding of the properties of outflows in relation to the properties of the AGN is still required.
 
AGN-driven outflows have been found to be multi-phase:   gas in all phases -- neutral atomic, molecular, and warm and hot ionised -- can be present.
Despite the large amount of energy released by the active black hole,  the cold component (i.e.\ atomic and molecular gas)  of the outflow often appears to be the most massive one, although observations are still limited to a relatively limited number of objects. Thus, these cold outflows likely have the largest impact in terms of kinetic energy and mass outflow rate (see e.g.\  \citealt{Feruglio10,Alatalo11,Dasyra12,Cicone14,Morganti05,Morganti13,Morganti15,Herrera19}),  with the possible exception of extremely powerful AGN (see \citealt{Fiore17,Brusa18}).

Outflows appear to be more common and more prominent in specific phases of the evolution of an AGN. The first phases of nuclear activity are particularly interesting in which the SMBH has just become active and  is still embedded in the interstellar medium (ISM) of the host galaxy. This  phase has been explored in young radio galaxies (see e.g.\ \citealt{Holt08,Holt09,Gupta06,Gereb15}), in highly obscured quasars (e.g.\ \citealt{Brusa18,Sun14}), and  in dust-obscured galaxies (DOGs; \citealt{Toba17,Fan18} and refs therein).  

In young radio galaxies, an outflow of cold gas can already start very close (a few tens of pc) to the active SMBH (\citealt{Schulz18}). This brings into question how cold gas can be involved in such energetic events so close to the AGN while also giving information on the relative lengths of the evolutionary time-scale of the AGN and the cooling time of the gas  (e.g.\  \citealt{Richings18}). It also raises  questions about possible interplay between these outflows close to the AGN and the infalling gas, which fuels the AGN.  

Several mechanisms have been suggested to  drive gas outflows. This is important  in the context of  the relevance of different types of AGN for feedback. 
Most commonly it is assumed that  outflows are driven by radiation pressure, or by a hot thermal wind  launched from the accretion disc that interacts with  the surrounding gaseous medium and extends to large scales \citep[e.g.][]{Faucher12,Zubovas12,Zubovas14}. Support for the latter mechanism comes from, for example, the correlation between the luminosity of the AGN and the properties of the outflows as seen in samples of different types of objects (e.g.\ \citealt{Fiore17,Fluetsch18}).

However, outflows can also be driven by radio plasma jets, and the number of known cases has been growing steadily, including low-power (e.g.\ \citealt{Alatalo11,Burillo14,Morganti15,Rodriguez17,Runnoe18,Husemann19a}) and  high-power radio AGN (e.g.\ \citealt{Nesvadba08,Holt09,Morganti05,Husemann19b}).
Numerical simulations are finding that radio plasma jets can actually couple strongly to the  clumpy ISM of the host galaxy \citep{Wagner12,Mukherjee16,Cielo18,Mukherjee18a,Mukherjee18b}. According to these numerical simulations, a clumpy ISM, instead of a smooth one, can make the impact of the jet much larger than previously considered. Due to the clumpiness of the gaseous medium, the progress of the jet can be temporarily halted when it hits a dense gas cloud and the jet  meanders through the ISM to find the path of minimum resistance and this creates a cocoon of shocked gas driving an outflow in all directions \citep{Wagner12,Mukherjee16,Mukherjee18a}.
The jet power, the distribution of the surrounding medium and the orientation at which the jet enters the medium are important parameters that determine the final impact  of such jet-ISM interactions \citep{Mukherjee18a}.

In this paper, we study the molecular gas in \pks, an object where the AGN is in the crucial early phases of  its evolution, while still being embedded in a dense ISM. 
\pks\ is one of the closest ($z=0.1525$) examples of a young, radio-loud quasar\footnote{The cosmology adopted in this paper assumes a flat Universe and the following parameters: $H_{\circ} = 70$ \kms\ Mpc$^{-1}$, $\Omega_\Lambda = 0.7$, $\Omega_{\rm M} = 0.3$. For the assumed redshift, 1 arcsec corresponds to 2.67 kpc.} ($P_{2.3 \rm{GHz}} = 2.7 \times 10^{25}$ W Hz$^{-1}$),  where the quasar nature is detected via broad emission lines in the near-IR, but is heavily obscured at optical wavelengths (\citealt{Bellamy03,Holt06}). 

In \pks\  a number of relevant processes are  happening. 
The AGN appears to be in the process of clearing its gas-rich surroundings in which it is enshrouded. A fast outflow is observed in the warm ionised gas \citep{Tadhunter01,Holt06,Batcheldor07} and an Ultra-Fast Outflow (UFO) has been revealed by X-ray observations (\citealt{Tombesi14}). \pks\ is also a small radio source of only about 300 pc in size, with an asymmetric core-jet structure (\citealt{Holt06}). This asymmetric structure suggests that relativistic orientation effects may play a role, or alternatively,  that a strong, asymmetric interaction occurs between the radio plasma and the clumpy surrounding medium.
\cite{Tadhunter01} and \cite{Holt06} presented a possible scenario where the newly born radio jet is fighting its way out the dense, rich medium of the merger remnant, accelerating the outflows and playing a key role in shedding its natal cocoon as  expected in the early phases of strong  feedback. 

At the same time, \pks\ has also undergone a recent major merger, as indicated by high-surface-brightness tidal tails (\citealt{Holt06,Batcheldor07}), bringing gas into the central regions of the host galaxy. Indeed,  observations suggest that a large amount of gas is surrounding the nucleus and the radio source, for example as deduced from  the high reddening along the line of sight to the quasar nucleus ($A_v > 4.9$, \citealt{Bellamy03,Holt06}). The presence of a rich medium has been confirmed by the detection of \HI\ 21-cm absorption (\citealt{Morganti01,Holt06}).
A young stellar population (50--250 Myr), likely resulting from gas accumulation from the recent merger, is also observed (\citealt{Tadhunter01,Holt06}). Further evidence for star formation is provided by its unusually strong far-infrared  emission, which leads to its classification as an ultraluminous infrared galaxy with $L_{\rm IR} = 1.6 \times 10^{12}$ \lsun.

Despite the high optical luminosity of the AGN and the powerful radio jet, the  kinetic energy associated with the warm, ionised gas outflow is only a tiny fraction of the Eddington luminosity of \pks, and this  outflow is  currently not capable of removing the gas from the bulge of the host galaxy, as required by  feedback models (\citealt{Holt06}).  Following what was found in other objects (e.g.\ \citealt{Feruglio10, Alatalo11,Cicone14,Morganti15}), much of the outflow may be  tied up in the cooler phases of the interstellar medium, in particular the molecular gas. 

Clearly, \pks\ represents an excellent object for studying how the energy released by the AGN couples to the rich surrounding medium in a stage of evolution of the AGN when feedback effects should be particularly prominent.
Here we present new \coOne\ and \coThree\ observations, as well as 1- and 3-mm continuum observations obtained with the Atacama Large Millimeter/submillimeter Array (ALMA).  We combine these data with new VLBI (LBA) 2.3-GHz images to shed light on the nature of the radio source and its role in the AGN-ISM interaction.

\begin{table*} 
\caption{Parameters of the ALMA and VLBI data cubes and images.
}
\begin{center}
\begin{tabular}{lcr@{ $\times$ }lccccc} 
\hline\hline 
   & Frequency & \multicolumn{3}{c}{Beam \& PA} &Velocity Res.  & Noise     & Description\\
   &  (GHz)   & \multicolumn{2}{c}{(arcsec) }  & (degree)& (\kms)          & (\mJybeam) & \\
\hline
\coOne\   &        &  0.167&0.126 & 9.5    & 18  & 0.29  & low resolution \\
\coOne\   &        &  0.090&0.055 & --2.9  & 120 & 0.085 & low velocity resolution \\
\coOne\   &        &  0.31&0.16   & --22.2 & 60  & 0.19  & cube matching \coThree\  \\
\coThree\ &        &  0.31&0.16  & --22.2  & 60  & 0.24 &        \\
Continuum &  100   &  0.047&0.026 & --10.5  &  --   & 0.085 \\
Continuum &  100   &  0.01&0.01   & 0      &   --  &     & super resolved  \\
Continuum &  300   &  0.29&0.13   & --22.5  &  --   & 0.25  \\
Continuum &  300   &  0.05&0.05   & 0      &  --   &    & super resolved\\
VLBI      & 2.3    &  0.0042&0.0012&  0  & -- &  0.8 \\
\hline
\end{tabular}
\end{center}
\label{tab:obs}
\end{table*}

\section{ALMA observations}
\label{sec:almaobservations}

\subsection{Observations of \coOne\ and 3-mm continuum}

The \coOne\ data  were obtained during Cycle 5 using Atacama Large Millimeter/submillimeter Array (ALMA) in configuration C43-10.
The observations were pointed at the nucleus of \pks, with a field of view (FoV) of $\sim$60". The observations were done in Band 3 making use of the correlator in Frequency Division Mode. The total bandwidth used was 1.875 GHz, corresponding to 5625 \kms\ using 1920 channels and giving a native velocity resolution of about 3.0 \kms, but in the subsequent data reduction channels were combined to make image cubes with a velocity resolution better matching the observed line widths (see below). The observations were done with 45 antennas, giving a $uv$ coverage with the shortest baseline being 35 k$\lambda$ and a maximum baseline of almost 5 M$\lambda$, in three observing sessions of 0.63 hr on-source each  (two  on Oct 8, 2017 and one on Oct 15, 2017), resulting in a total on-source time of 6804 sec.  In each observation session, we interleaved 1-minute scans on \pks\ with 20-second scans on the phase calibrator J1617--7717 .

The initial calibration was done in CASA (v5.1.1; \citealt{McMullin07}) using the  reduction scripts provided by the ALMA observatory.  The flux calibration was done using scans on the source J1617--5848. These calibrated $uv$ data were  exported to MIRIAD \citep{Sault95} which was used to perform additional bandpass- and self calibration which improved the quality of the images significantly. 
The continuum flux density of \pks\ at 100 GHz is about 370 mJy and most of this comes from the unresolved core of the source. The pipeline provided by the ALMA observatory, due to this relatively bright core,  resulted in data cubes with insufficient spectral dynamic range. To remedy this, we used the scans of the phase calibrator  J1617--7717 (which has a flux density of 2.9 Jy) that were interleaved with the scans on \pks\ to derive a time-variable bandpass calibration for all three \coOne\  observations. This greatly improved the bandpass calibration, and all final data cubes we produced are not limited in spectral dynamic range.
All further reduction steps (continuum subtraction in the image plane, imaging, cleaning) were also done in MIRIAD.

   \begin{figure}
   \centering
       \includegraphics[angle=0,width=7.5cm]{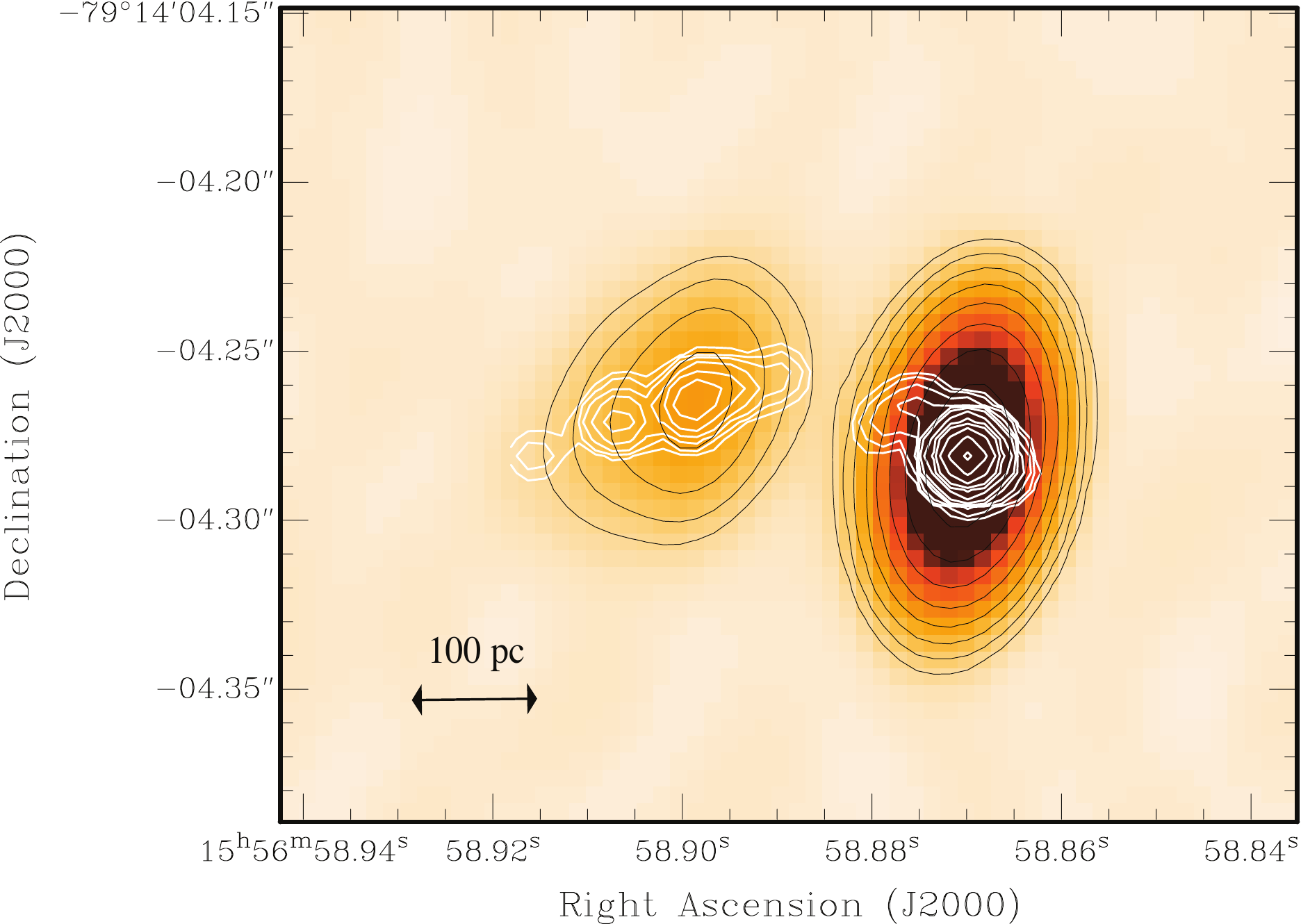}
   \caption{Continuum image obtained from the line-free channels of the 3-mm observations (black contours and grey-scale). The white contours show the structure of the same emission as obtained from the super-resolution image. Contour levels for standard image are 1.5, 3.0, 6.0, 12.0,... \mJybeam, for the super-resolved image 0.15, 0.3, 0.6, 1.2,... \mJybeam.}
              \label{fig:continuum}
    \end{figure}

   \begin{figure}
   \centering{
       \includegraphics[width=6.5cm,angle=-90]{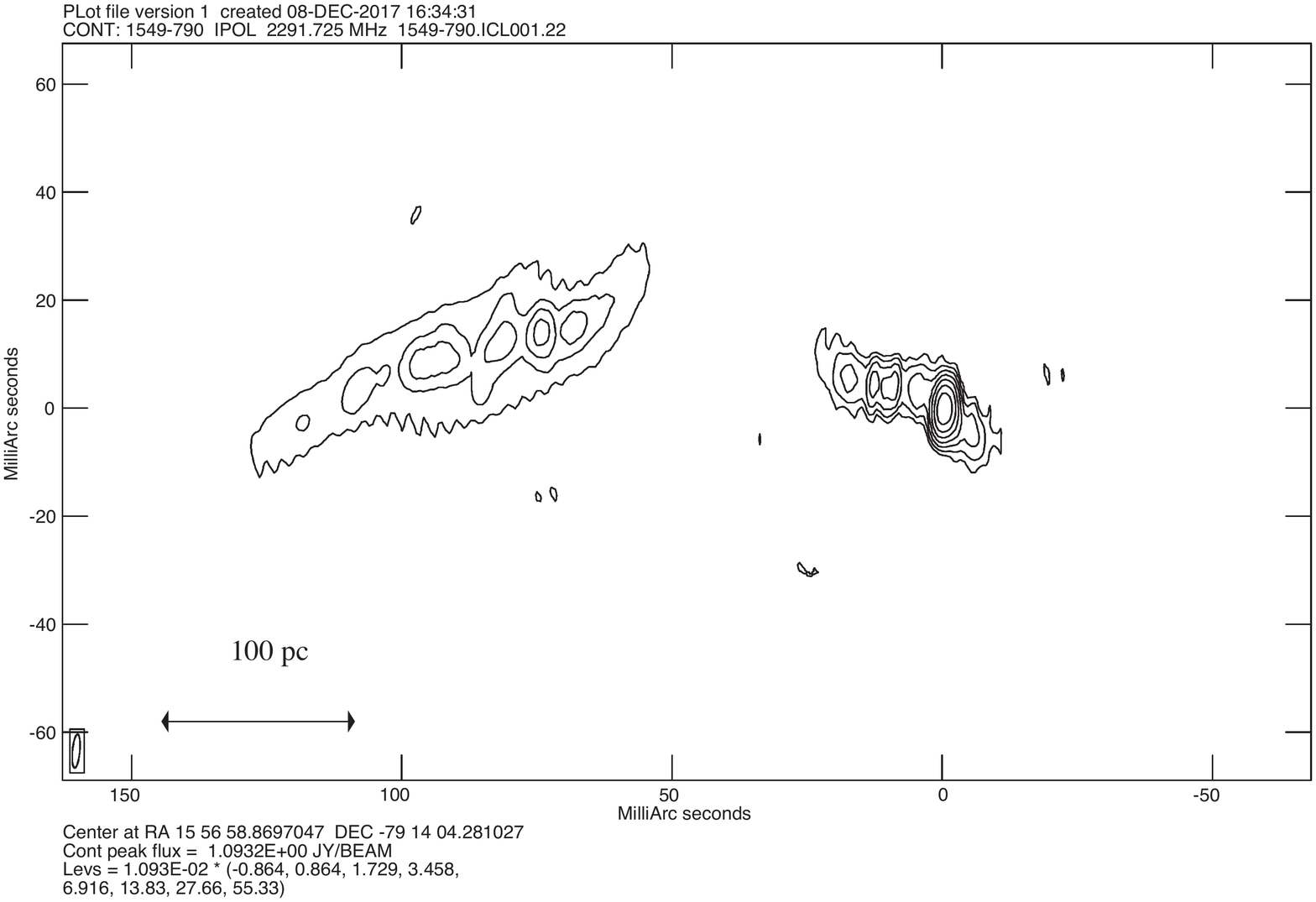}
}
   
   \caption{VLBI continuum image at 2.3 GHz. The contour levels are $0.011\times -0.864,0.864,1.729,3.458,6.916,13.83,27.66,55.33$ Jy beam$^{-1}$.}
              \label{fig:vlbicontinuum}
    \end{figure}

As listed in Table \ref{tab:obs}, a number of cubes were made, adopting various  weighting schemes and velocity resolutions in order to explore the optimum for imaging and for highlighting different structures of the distribution and kinematics of the \coOne. 
Given the low surface brightness of the \coOne\ emission, we  produced a data cube using natural weighting to obtain the lowest noise level, albeit at somewhat lower resolution (0\farcs09 $\times$ 0\farcs05 = 240 $\times$ 147 pc). This cube was used to study the inner regions of \pks. The velocity resolution of this cube is 120 \kms\ to match the large line widths of the \coOne\ close to the centre. To better image the large-scale \coOne, we also made a cube with lower spatial resolution by tapering the data. The resolution of this cube is  0\farcs167 $\times$  0\farcs126 (446 $\times$ 336 pc) and has a velocity resolution of  18 \kms. All cubes were de-redshifted assuming a redshift of $z = 0.1525$. 

The 100-GHz continuum image was obtained by imaging the data, after self-calibration,  at full spatial resolution using uniform weighting with a resulting beam of 0\farcs047 $\times$ 0\farcs026 (125 $\times$ 69 pc). The noise of the continuum image is 85 \muJybeam. The peak in the continuum image is 350 mJy at the location of the core, so the dynamic range is about 1:4000. The total extent of the radio source is about 0\farcs2 (about 500 pc) with an inner jet  of 0\farcs06 (about 140 pc; see Fig.\ \ref{fig:continuum}).

Given the high quality of the data, the large number of antennas used, the resulting excellent $uv$ coverage, and the strength of the central continuum point source, it turned out that the source model derived in the selfcalibration (using the clean method) contained information on smaller scales than the nominal resolution of the observations. In Fig.\ \ref{fig:continuum} we show both the nominal continuum image and this model image (where we have smoothed the model components for presentation purposes with a Gaussian   with a resolution of 0\farcs01 corresponding to 27 pc). This very high resolution image compares very well with  the 2.3-GHz VLBI image we discuss below (Fig.\ \ref{fig:vlbicontinuum}), lending support to its fidelity.

   \begin{figure*}
   \centering
\includegraphics[width=8cm,angle=0]{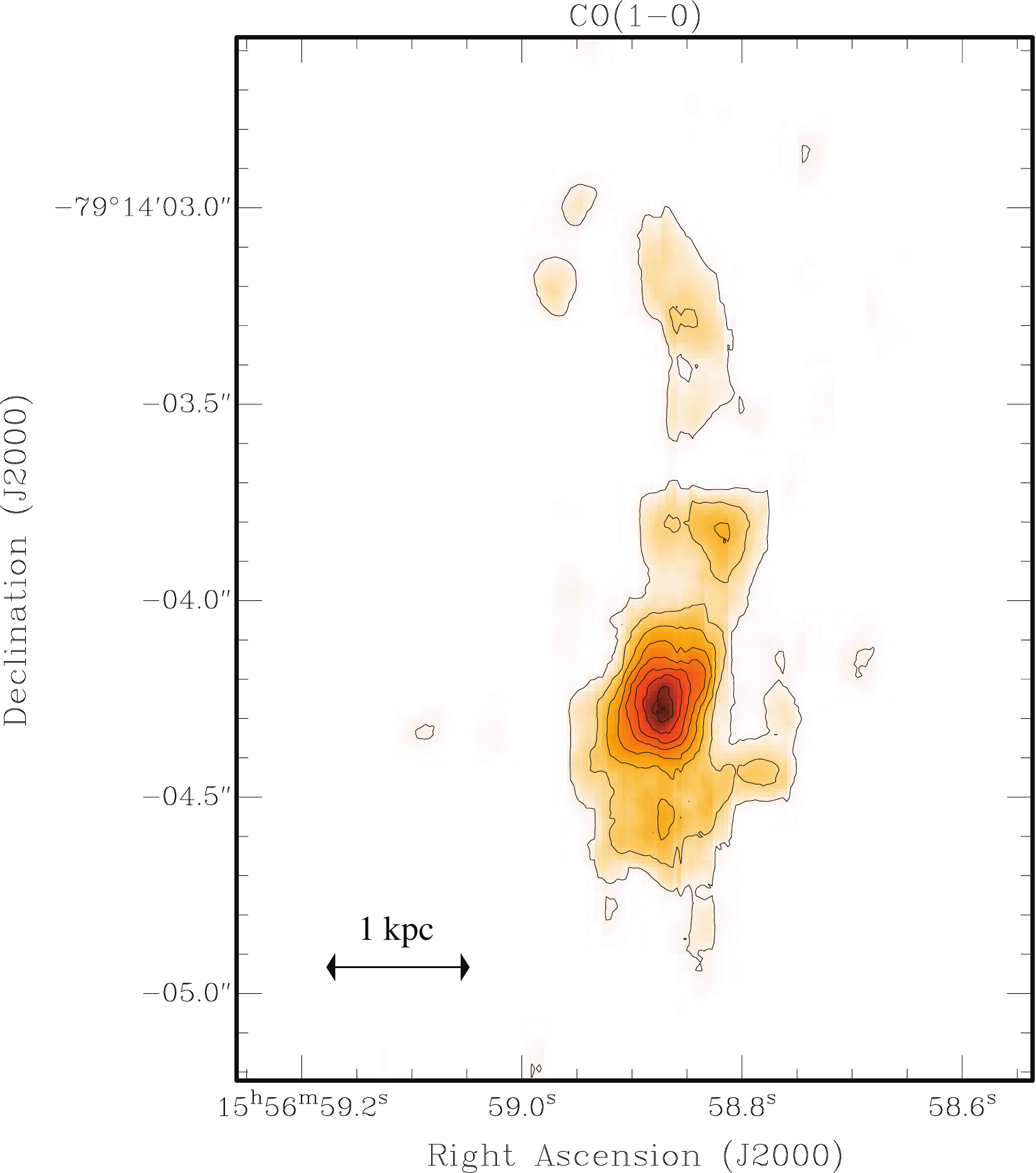}\hskip1cm
\includegraphics[width=8cm,angle=0]{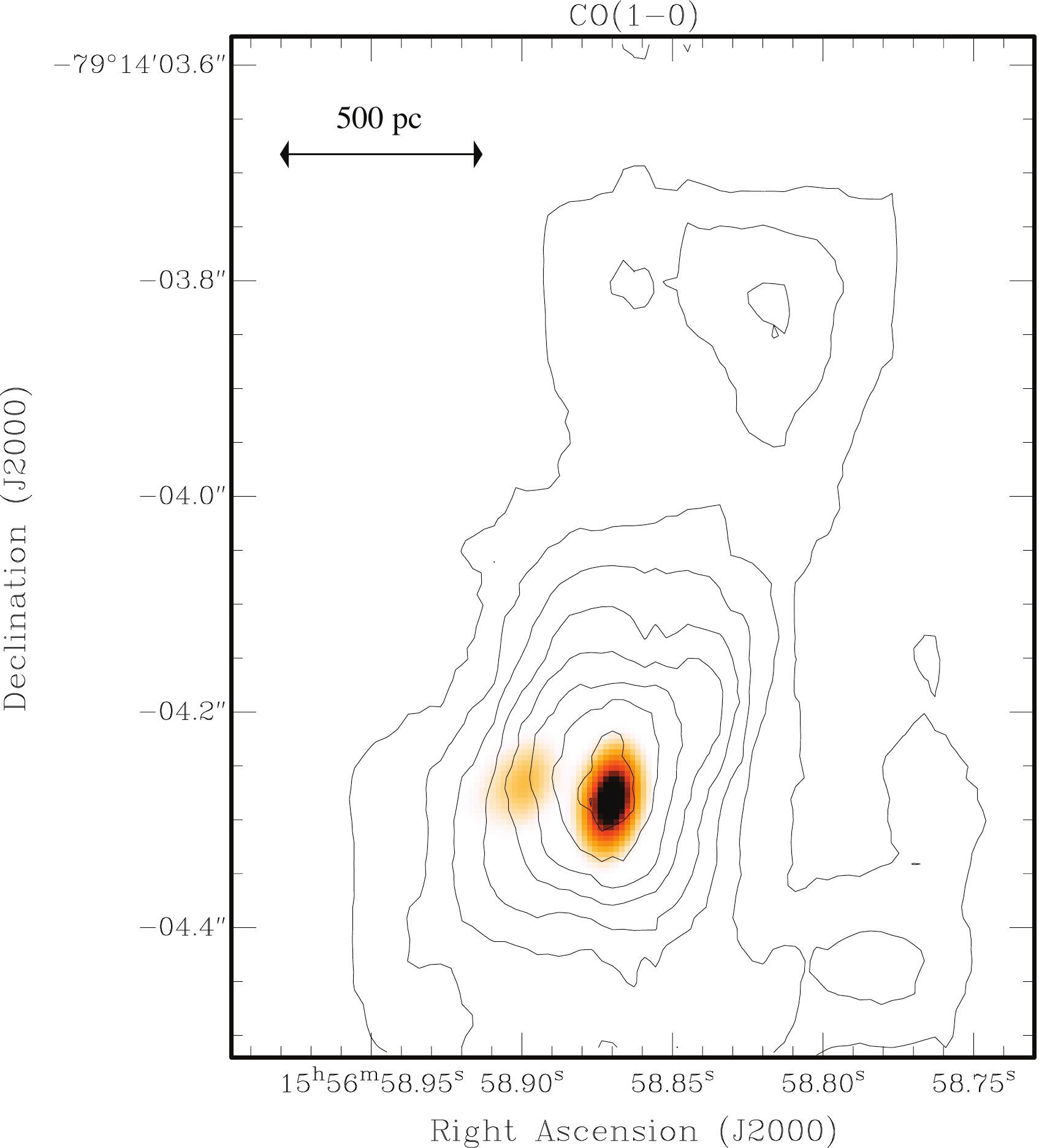}
   \caption{Total intensity of the \coOne\ obtained from the cube with a resolution of 0\farcs17 $\times$ 0\farcs13, corresponding to about 400 pc. This low-resolution allows to highlight the extent of the N-S tail of \coOne. In the right hand panel the continuum image is shown for reference (see details in Sec. \ref{sec:results_continuum}). Contour levels are 0.04, 0.08, 0.12, ... \mJybeam \kms.}
              \label{fig:IntensityCO}%
    \end{figure*}

   \begin{figure*}
   \centering
    \includegraphics[width=8cm,angle=0]{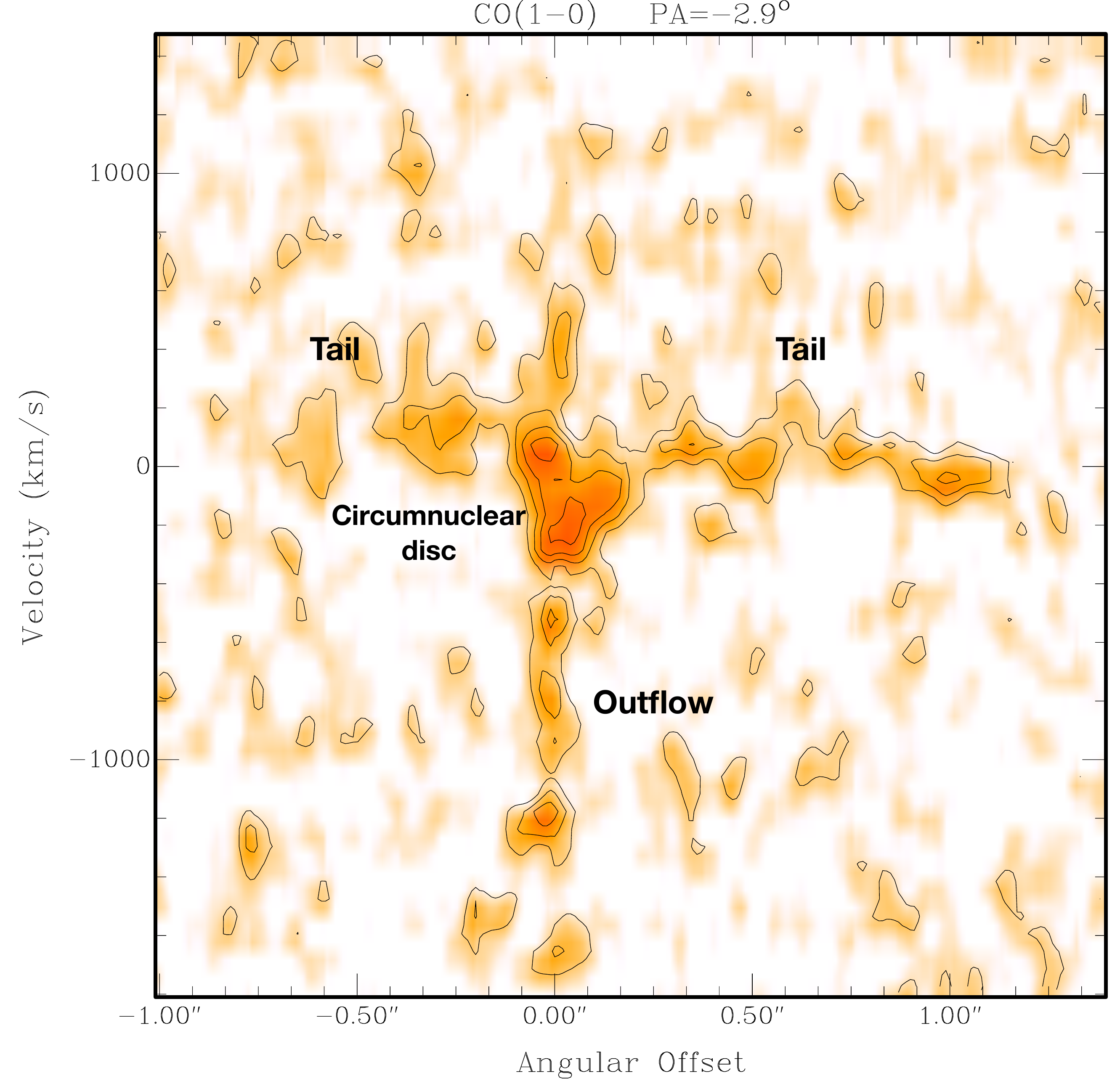}
    \hskip1cm
    \includegraphics[width=8.0cm,angle=0]{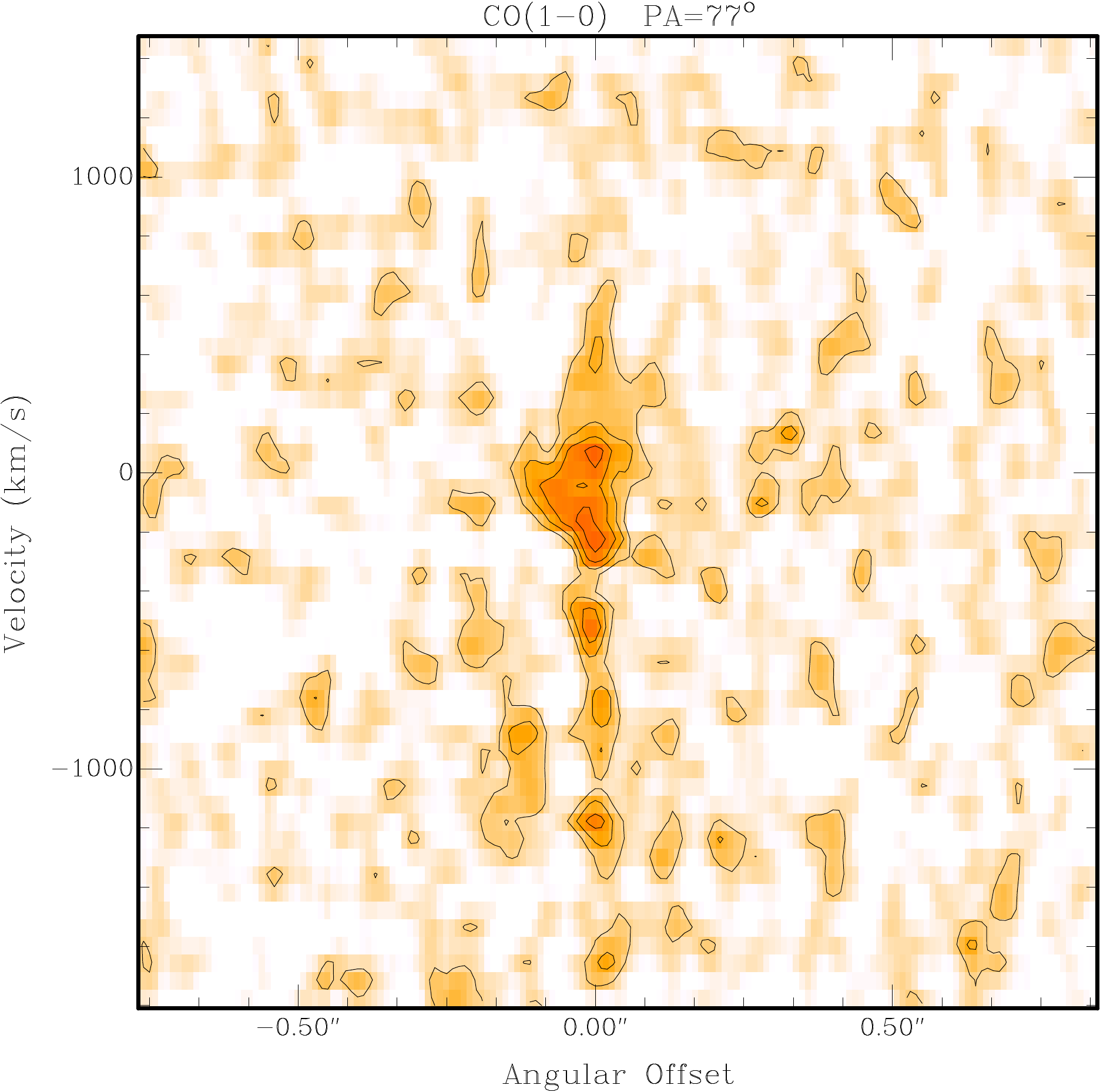}
   \caption{{\sl left: } Position-velocity plot of the \coOne\ taken along PA = --2\fdg9, i.e.\ along the direction of the inner parts of tails and almost perpendicular to the radio jet. 
   The data cube used has a resolution of 0\farcs09 $\times$ 0\farcs055 and a velocity resolution of 120 \kms. Negative position offsets are south of the core. Contour levels are 0.12 (1.5 $\sigma$), 0.24, 0.36 and 0.48 \mJybeam. {\sl right:} Position-velocity plot of the \coOne\ centred in the core in the direction of the radio jet (PA = 77$^\circ$). negative offsets are E of the core. Contour levels are 0.12 (1.5 $\sigma$), 0.24,  0.36 and 0.48  \mJybeam.}
              \label{fig:kpv}%
    \end{figure*}

\subsection{Observations of \coThree\ and 1-mm continuum}

The \coThree\  data were also obtained during Cycle 5, in a single observing session on Sep 20, 2018 using ALMA in configuration C43-5. The total on-source observing time was 0.75 hr. The observation made use of 45 antennas giving a $uv$ coverage with the shortest baseline being 15 k$\lambda$ and a maximum baseline of 1 M$\lambda$. The flux calibration was done using observations of J1427--4206.
 The observations were done in Band 7, again making use of the correlator in Frequency Division Mode. The spectral setup used was the same at that for the \coOne\ observations, giving a velocity resolution (due to the different observing frequency) of  about 1.0 \kms\ (de-redshifted), but also here in the later data reduction channels were combined to make image cubes with a velocity resolution better matching the observed line widths. 

The data reduction followed a very similar path as described for the \coOne, but given the lower continuum flux density ($\sim$60 mJy), additional bandpass calibration was not needed.
The aim of the observations  was to study to what extent the molecular gas is effected by the AGN by comparing  the  \coOne\ and \coThree\ emission and kinematics. Given the limited bandwidth  of the \coThree\ data in \kms\ (a third of that of the \coOne\ observations),  this is only possible for the for the larger, brighter CO  structures with velocities within $\sim$500 \kms\ from the systemic velocity, but not  for the gas with the most extreme outflow velocities detected in \coOne\ (see below). 
As summarised in Table \ref{tab:obs}, we obtained a cube with spatial  resolution of 0\farcs31 $\times$ 0\farcs16  (0.83 $\times$ 0.43 kpc). The noise of the cube used is 0.24 \mJybeam (for a velocity resolution of 60 \kms). A  cube with matching spatial and velocity resolution was made from the \coOne\ observations.

The continuum was imaged at full resolution (0\farcs29 $\times$ 0\farcs13 = 780 $\times$ 347 pc). The noise in the continuum image is 0.25 \mJybeam. Also for these observations we were able to make a super-resolved continuum image with in this case a resolution of 0\farcs05. Because of the  lower resolution of these observations, the continuum images give information on the spectral index of the continuum emission, but does not add information on the structure of the continuum source.

\section{VLBI observations}
\label{sec:vlbiobservations}

\pks\ was observed at 2.3 GHz in June 2007 (project V235) using the Australian Long Baseline Array (LBA) with in addition the Hartebeesthoek 26-m antenna (South Africa). Stations involved were the ATCA tied array, Mopra, Parkes, Hobart 26-m, Ceduna, and Tidbinbilla 34-m in Australia, and the Hartebeesthoek 26-m antenna in South Africa. There is a large gap in $(u,v)$ coverage between the baselines to Hartebeesthoek and the baselines between the Australian telescopes. The total observation covered 12 hours, with almost 11 hours integration on \pks\ at most stations; the source was visible at Hartebeesthoek only for the last 3.5 hours, and Tidbinbilla observed for just over three hours due to its limited availability for scheduling. The recorded bandwidth was 64 MHz in total, centred on a sky frequency of 2.3 GHz. Ceduna and Hobart recorded only the lower 32 MHz, limited by the recording capability at the time. All stations recorded dual polarisation data except for Tidbinbilla, which has only single polarisation (right-hand circular). The data were correlated using the DiFX software correlator at Swinburne University (\citealt{Deller07}).

Amplitude scaling based on nominal System Equivalant Flux Densities (SEFDs) for each antenna was applied at correlation, as was standard procedure for the LBA correlator at that time. Post-correlation, these a priori corrections were undone and scaling based on the measured system temperatures was applied where available, in this case only for Hobart, Ceduna and Tidbinbilla. This calibration, along with fringe-fitting, initial off-source and band-edge flagging, and bandpass calibration, was performed using standard tasks in AIPS. The data were then independently imaged in both AIPS and Difmap.

The resulting synthesised beam for the full dataset has FWHM $4.2 \times 1.2$ mas ($11.2\times5.3$ pc). Initially, several iterations of  {\sc CLEAN} and phase-only self-calibration were done, followed by an overall amplitude scaling using the latest {\sc CLEAN} component model to scale the visibilities. This resulted in typical amplitude corrections in the range 5--20\% for each antenna and 16-MHz sub-band. Further iterations of {\sc CLEAN}, phase-only self-calibration, and phase and amplitude self-calibration with a 30-minute solution interval, were used to refine the image. The resulting residual image has an RMS noise level of 0.8 mJy/beam, compared to the peak flux density of 0.76 Jy/beam, giving a dynamic range of approximately 1000:1 in the full resolution image with uniform weighting. The resulting image is shown in Fig.\ \ref{fig:vlbicontinuum}.

\section{Results: the molecular gas}
\label{sec:results_co}

The ALMA observations clearly reveal the complex distribution and kinematics of the molecular gas in \pks\ on several scales.  On the largest scale (i.e.\ few kpc), the emission appears to form extended tails in the north-south direction. In the inner few hundred parsec we see evidence for a circumnuclear disc.  In addition, in \coOne\ we detect a very broad ($\sim$2300 \kms) component at the position of the AGN. We do not  detect this component in \coThree, but this is quite likely due to the limited bandwidth of the \coThree\ data. 
We  first describe the results for the two transitions separately, followed by a discussion of the line ratios.

\subsection{Overall distribution and kinematics of the CO(1-0)}

The total intensity of the \coOne\ is shown in  Fig.\  \ref{fig:IntensityCO} and the overall kinematics is illustrated in Fig.\ \ref{fig:kpv}. We detect three main features in \coOne. On the largest scales of several kpc, we see that the \coOne\  extends about 1\farcs7 (4.5 kpc) in the N-S direction in what appear to be two tails of gas. This orientation is perpendicular, in projection, to that of the radio jet, which emanates from the core more or less towards the east.
The overall morphology of the CO is very similar to that seen for the ionised gas as observed with the Hubble Space Telescope  \citep[HST;][]{Batcheldor07} and for the stellar distribution with the VLT \citep{Holt06}, which also show N-S tail-like structures. The ALMA observations show that these tails go all the way to the very central regions. The  zoom-in of the central region  (Fig.\  \ref{fig:IntensityCO}) shows that the \coOne\ strongly peaks at the centre, at the location of the radio core. As is shown below, this is at least to some extent the effect of  high excitation of the central CO gas and not only of a very high central concentration of the gas.   At the location of the jet east of the core, no anomalous feature is seen in  \coOne.

   \begin{figure}
   \centering
    \includegraphics[width=8cm,angle=0]{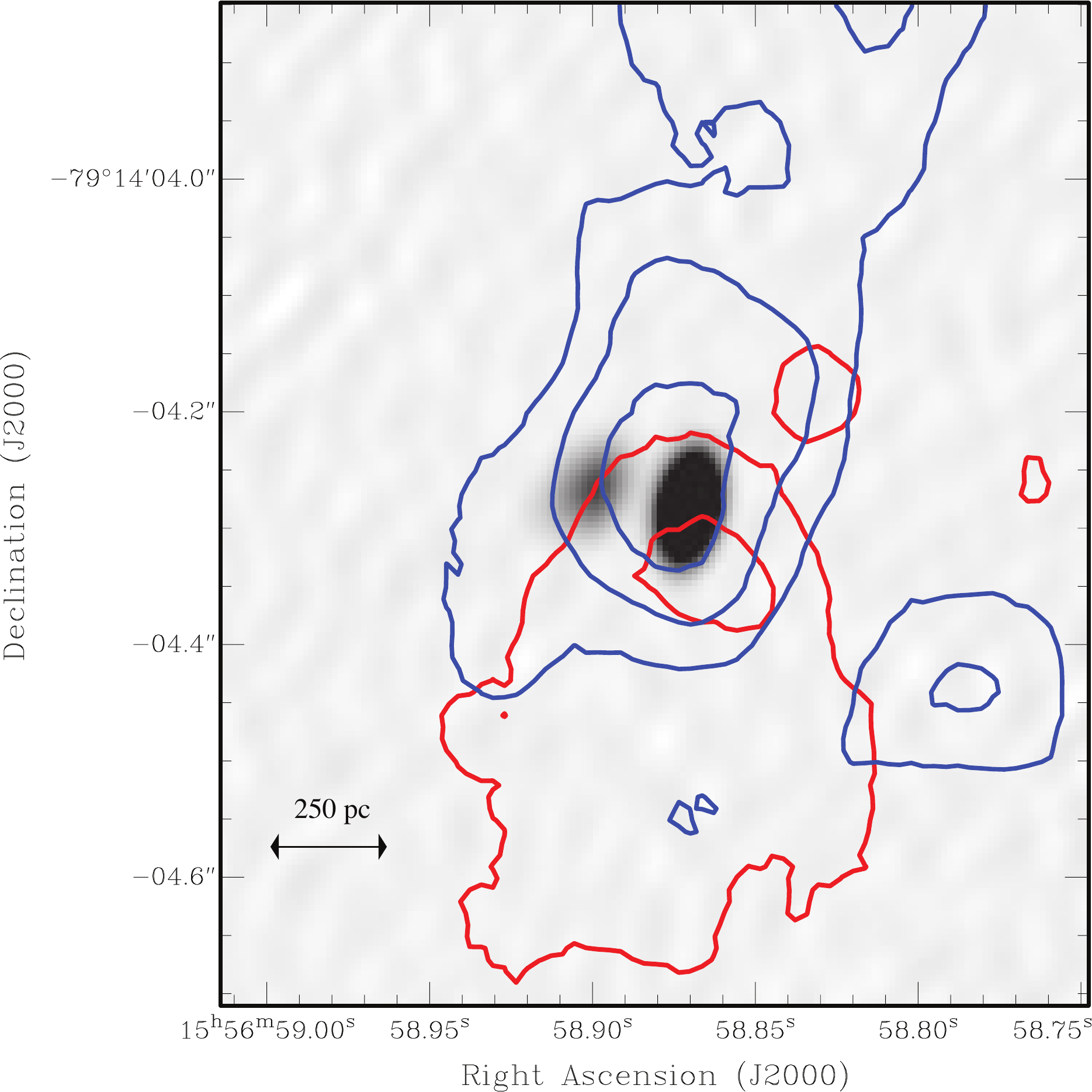}
   \caption{Total blue- and redshifted \coOne\ emission of the  circumnuclear disc  superposed on the 100-GHz continuum image, illustrating the N-S velocity gradient (see text for details).  Contour levels are 0.04, 0.08 and 0.12 \mJybeam\ \kms.}
              \label{fig:bluered}%
    \end{figure}

The position-velocity plot in the left panel of Fig.\ \ref{fig:kpv}  shows the kinematics of the \coOne\ in the N-S direction (PA --2\fdg9, centred on the core and aligned with the inner parts of the large-scale tails) as seen in the naturally weighted cube. The velocity resolution of the data cube used here is 120 \kms\ to highlight the broad component in the centre (see below).  A few things can be noted from this figure. In the first place that the velocity gradient over the large-scale tails is very small. This is somewhat surprising because the linear morphology of the tails suggests that we are seeing the large-scale gas distribution of \pks\ edge-on so that one should fully detect any  rotation of the large-scale gas.  The small velocity gradient observed may suggest that the kinematics of the large-scale gas is dominated by radial motions (which will mostly be in the plane of the sky). If this is the case, the data suggest that there could be  large-scale inflow of gas towards the central regions. This would be consistent with the idea that \pks\ is an ongoing merger with strong star formation fed by the accretion of gas.

In contrast to what is seen on the largest scales, in the inner region ($\sim$0\farcs5 = 1.3 kpc) there appears to be a distinct kinematical component with  a total  velocity width  of  $\sim$500 \kms. The kinematics suggests that on this scale,  gas has settled, or is settling, in a rotating circumnuclear disc with a diameter of just over 1 kpc. This circumnuclear disc is even better visible in the \coThree\ data (see below). 
The kinematics of the gas in the inner regions is further illustrated in Fig.\ \ref{fig:bluered} where we show the the blue- and redshifted \coOne\ emission of the circumnuclear disc. We have isolated these components  by integrating the data over  the intervals \hbox{--350} \kms\ to 0 \kms\  and 0 \kms\ to +350 \kms\ respectively. Figure \ref{fig:bluered} shows that the gas from the circumnuclear disc is extended N-S with the velocity gradient also in that direction. 

\begin{figure}
   \centering
       \includegraphics[width=8cm,angle=0]{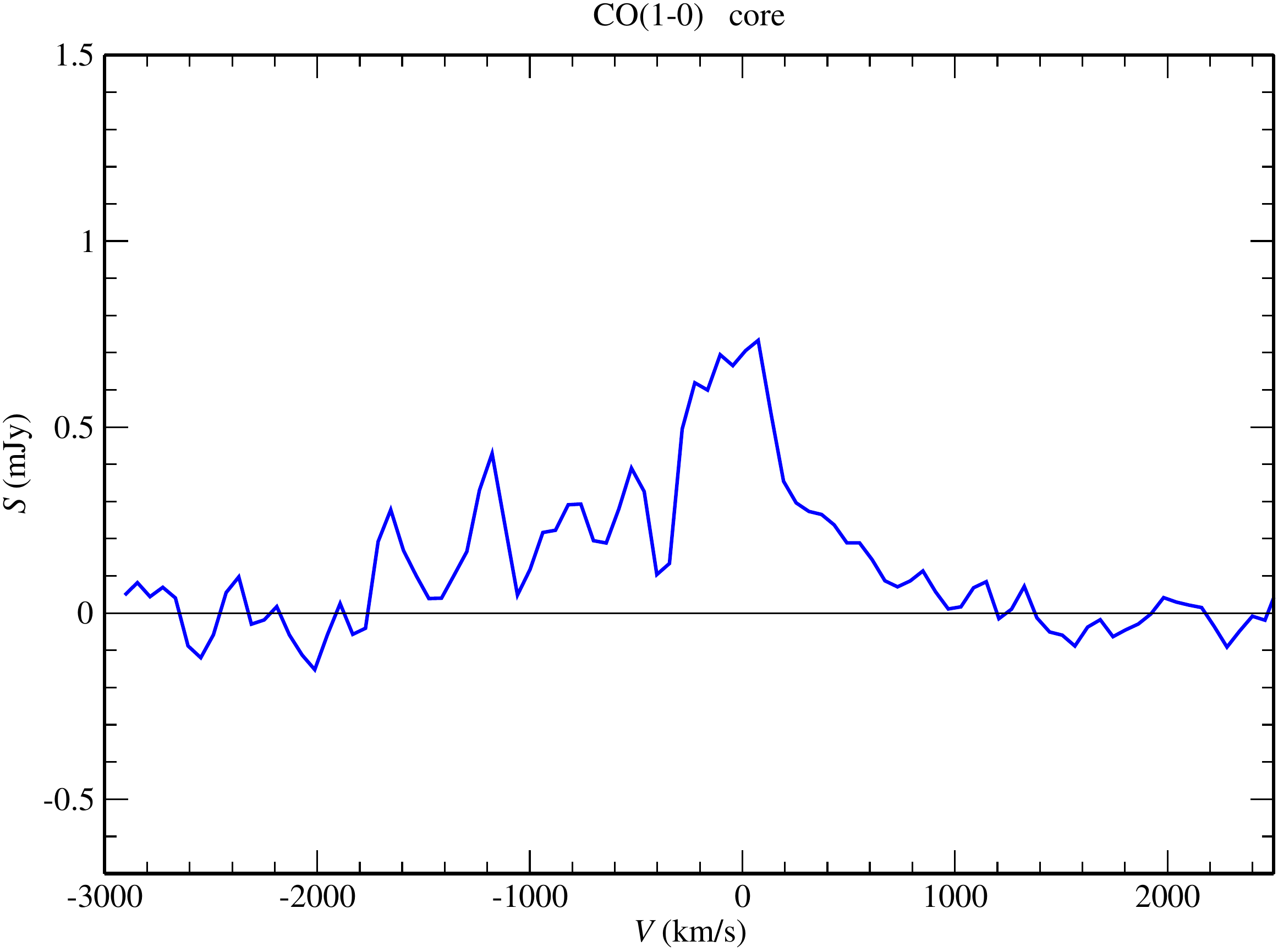}
   \caption{Profile of the CO(1-0) emission at the location of the core of \pks}
              \label{fig:profile}
    \end{figure}

Most interestingly, in the very central region, at the location of the core, \coOne\ is detected over a very large velocity range, with a total width of about 2300 \kms.  A spectrum taken at the position of the core is shown in Fig.\ \ref{fig:profile}. This very broad profile shows both a blue-shifted component (up to $\sim$1800 \kms\ from the systemic velocity) and a narrower red-shifted wing ($\sim$500 \kms\ from systemic). In the optical, the \OIII\ emission line shows a similarly broad, and similarly blue-shifted profile (\citealt{Holt06}; Santoro et al.\ in prep.).  Most likely, the broad emission near the core is evidence that the AGN is affecting the gas in its vicinity inducing large turbulent motions and driving a gas outflow. This broad component is spatially unresolved in our data ($r < 120$ pc) and is only detected at the location of the core. No indications for large, anomalous velocities are seen at the location of the  jet, about 0\farcs1 east from the core, even after spatial tapering of the data to enhance faint, extended emission. 

\HI\ absorption is known to be present in \pks\  against the entire radio structure, including the few-hundred-pc sized jet (see \citealt{Morganti01,Holt06}). The \HI\ absorption profile is much narrower (80 \kms) than the broad CO profile, with velocities covering the blueshifted edge of the velocity range of the circumnuclear disc and the large-scale tails. The small width of the \HI\ profile suggests that the \HI\ is likely located at larger radii and not affected by the interaction with the jet and is possibly associated with the larger-scale CO tails.

\subsection{The CO(3-2) distribution and kinematics}
\label{sec:results_co2}

\begin{figure}
   \centering
    \includegraphics[width=8.5cm,angle=0]{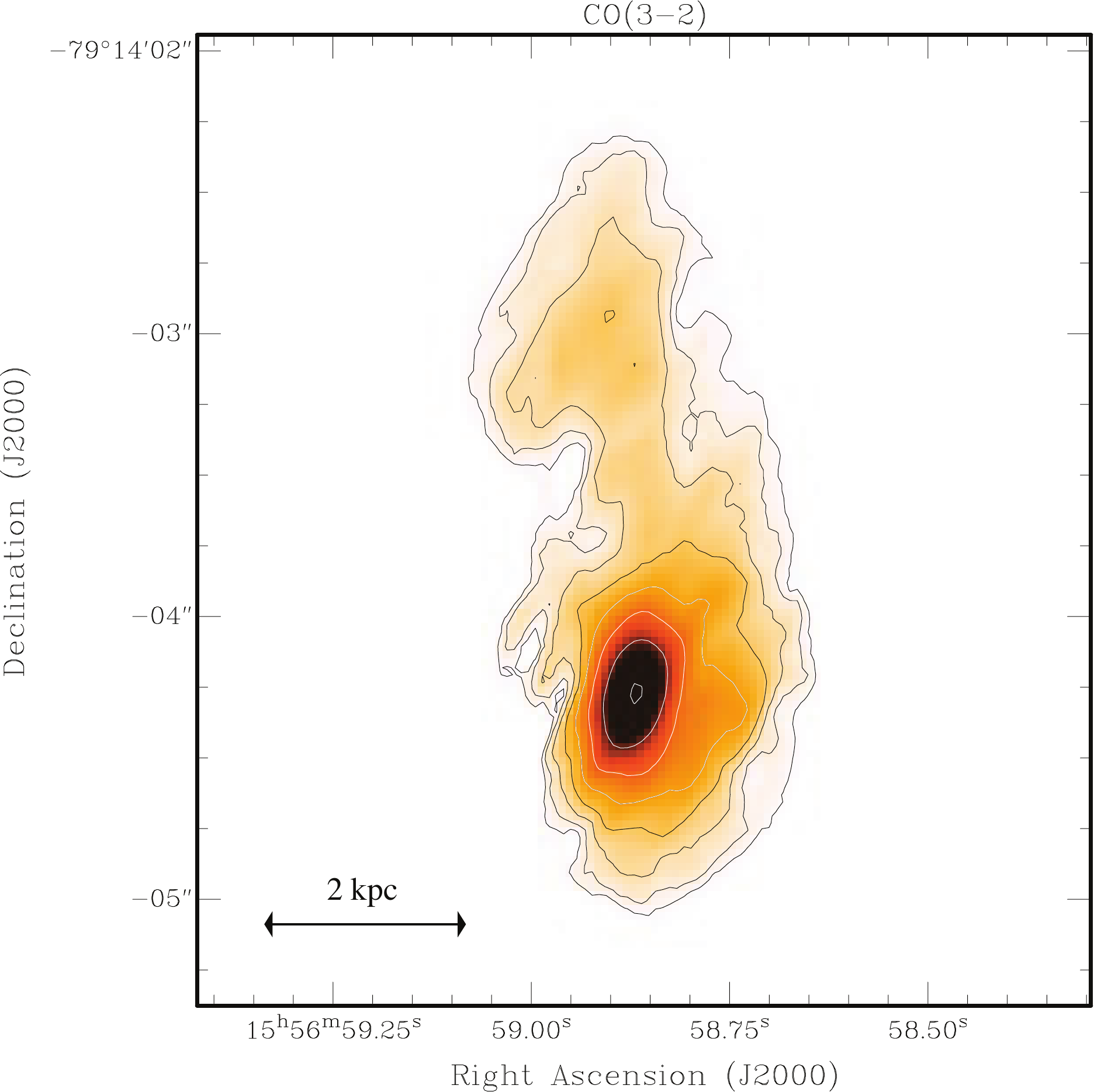}
   \caption{Total intensity of the \coThree. The resolution is 0\farcs31 $\times$ 0\farcs16 (820  $\times$  427 pc). Contour levels are 0.04, 0.08, 0.16,... \mJybeam\ \kms.}
              \label{fig:tot3-2}%
    \end{figure}

In Fig.\ \ref{fig:tot3-2} we show the integrated \coThree\ emission with a resolution of 0\farcs31 $\times$ 0\farcs16 (820 $\times$ 427 pc).
As expected, overall the CO(3-2) follows  the distribution  of the CO(1-0) quite well, although the contrast between the central regions and the outer tails is much larger,  with the central regions being relatively much brighter in \coThree.  This is very similar to what is seen, for example, in IC 5063 where the contrast between the inner regions affected by the radio jet and the outer, quiescent disc is larger in the higher CO transitions, evidencing the impact of the AGN on the gas conditions \citep{Oosterloo17}. 

\begin{figure}
   \centering
    \includegraphics[width=8.5cm,angle=0]{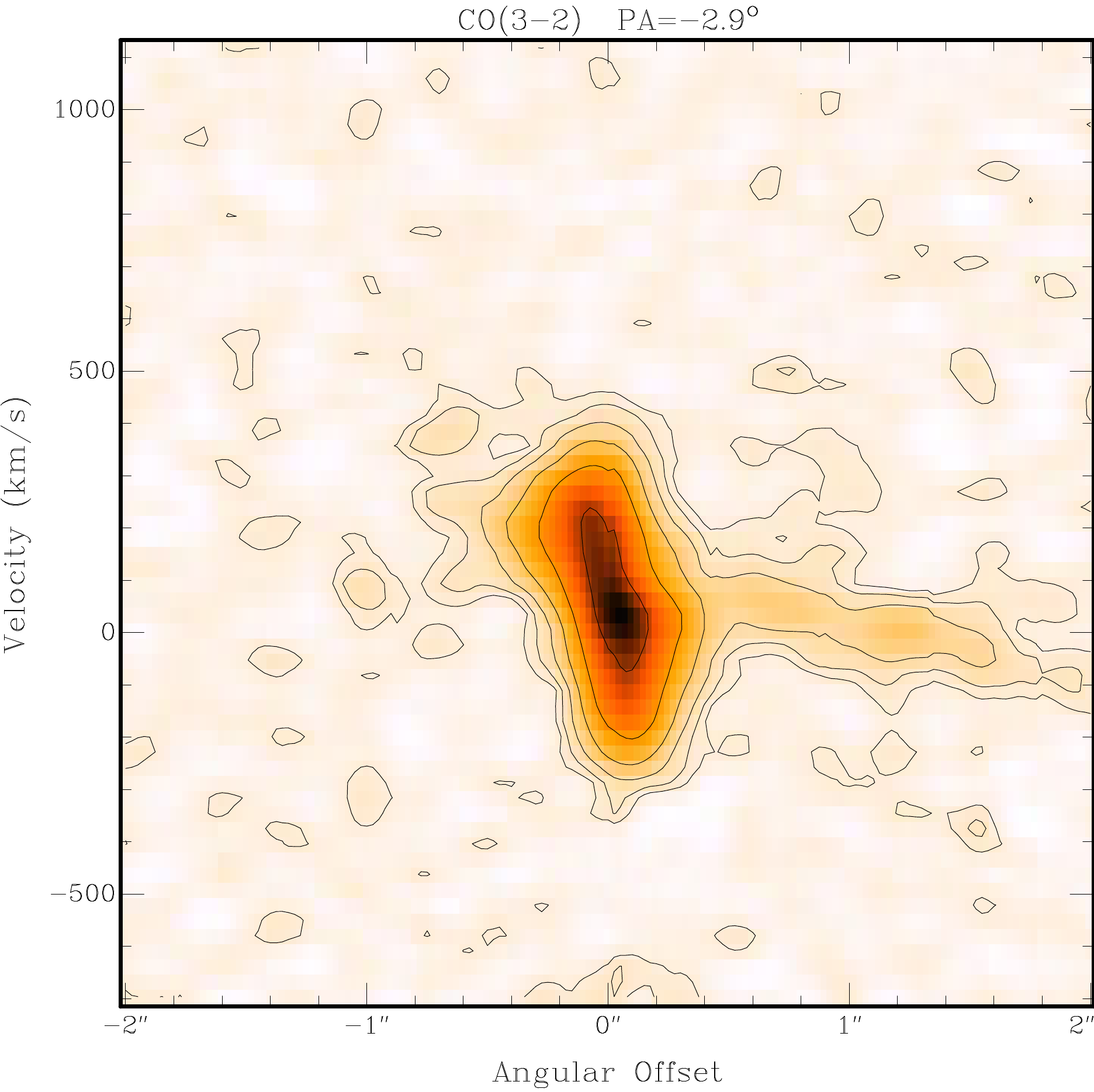}
   \caption{Position-velocity plot along the same direction as in Fig. \ref{fig:kpv} but from the \coThree\ data. The spatial resolution of this cube is 0\farcs31 $\times$ 0\farcs16 and the velocity resolution is 60 \kms. Contour levels are 0.36 (1.5 $\sigma$), 0.72, 1.44, 2.88,... \mJybeam.}
              \label{fig:kpv32}%
    \end{figure}

Figure \ref{fig:kpv32}  shows the kinematics of the \coThree\ along the same direction as shown for the \coOne\ in Fig.\ \ref{fig:kpv}. The central circumnuclear disc shows up very clearly in this transition, while the figure also shows the clear difference in brightness between the inner regions and the outer tails, underlining they are two distinct components. The broad component near the core seen in \coOne\ is not visible in \coThree. This may be a real effect, but it might also be the consequence of the relatively narrow observing band  used for the \coThree\ ($\sim$1800 \kms) so that the broad component was subtracted away in the continuum subtraction, given that the broad component seen in \coOne\ would extend to outside the observing band of the \coThree\ observations.

  
\subsection{Line ratios}
\label{sec:lineratios}

The very broad profile seen in the centre of \pks, and the large contrast in \coThree\ between the bright inner disc and the outer tails, suggest that the energy released by the AGN has an impact on the ISM surrounding it, both on the kinematics of the gas, and on the physical conditions. A way of illustrating this is to look at the line ratio \coThree/\coOne\ because this gives indications about the excitation conditions of the molecular gas.

In order to be able to do this, data cubes were made with matching spatial and velocity resolutions for both transitions. Given the higher resolution of the \coOne\ observations, a separate data cube for this transition was made by tapering the data so that the spatial resolution matches that of the \coThree\ cube discussed above.

\begin{figure*}
   \centering
    \includegraphics[width=10cm,angle=0]{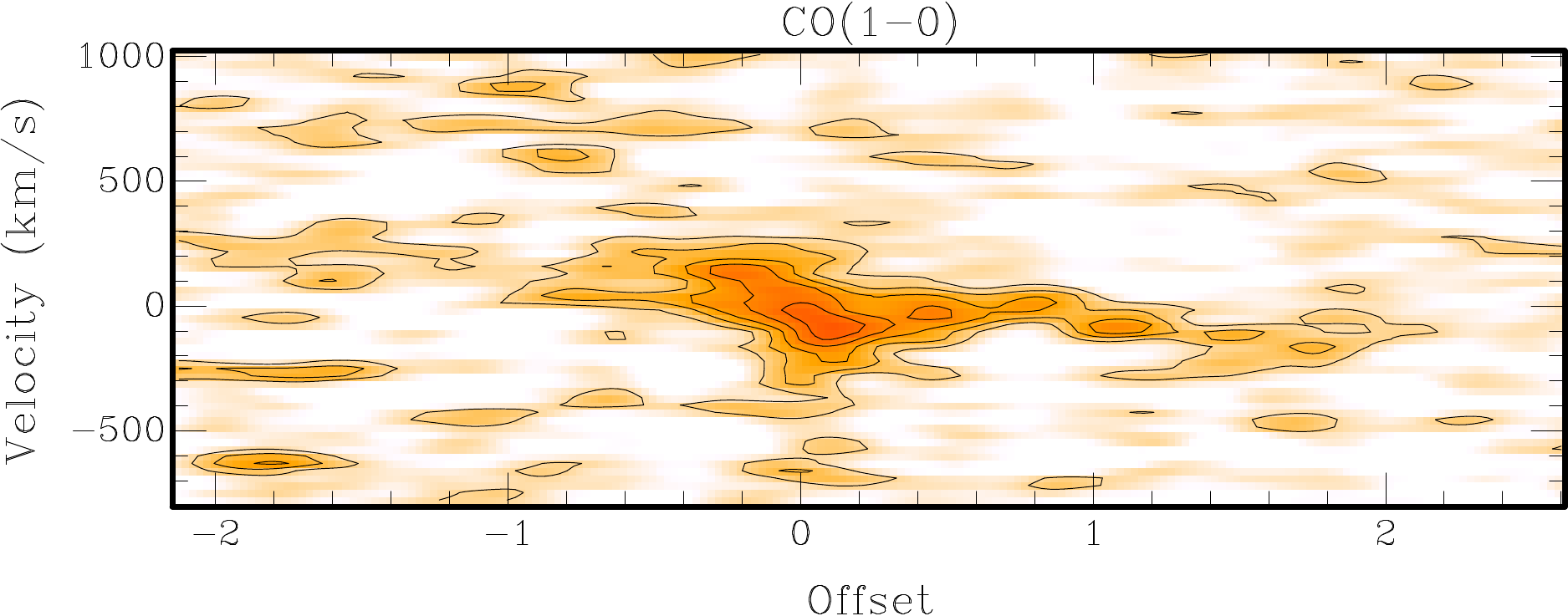}
    \includegraphics[width=10cm,angle=0]{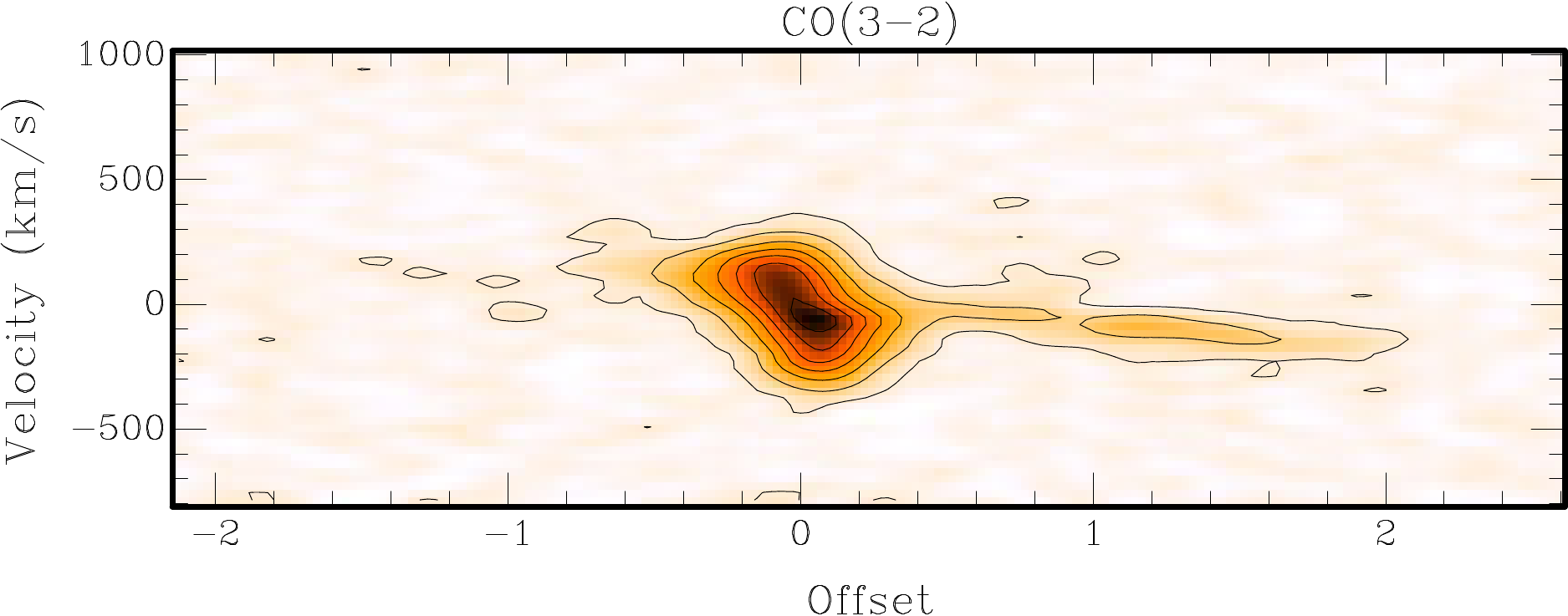} 
  \includegraphics[width=12cm,angle=0]{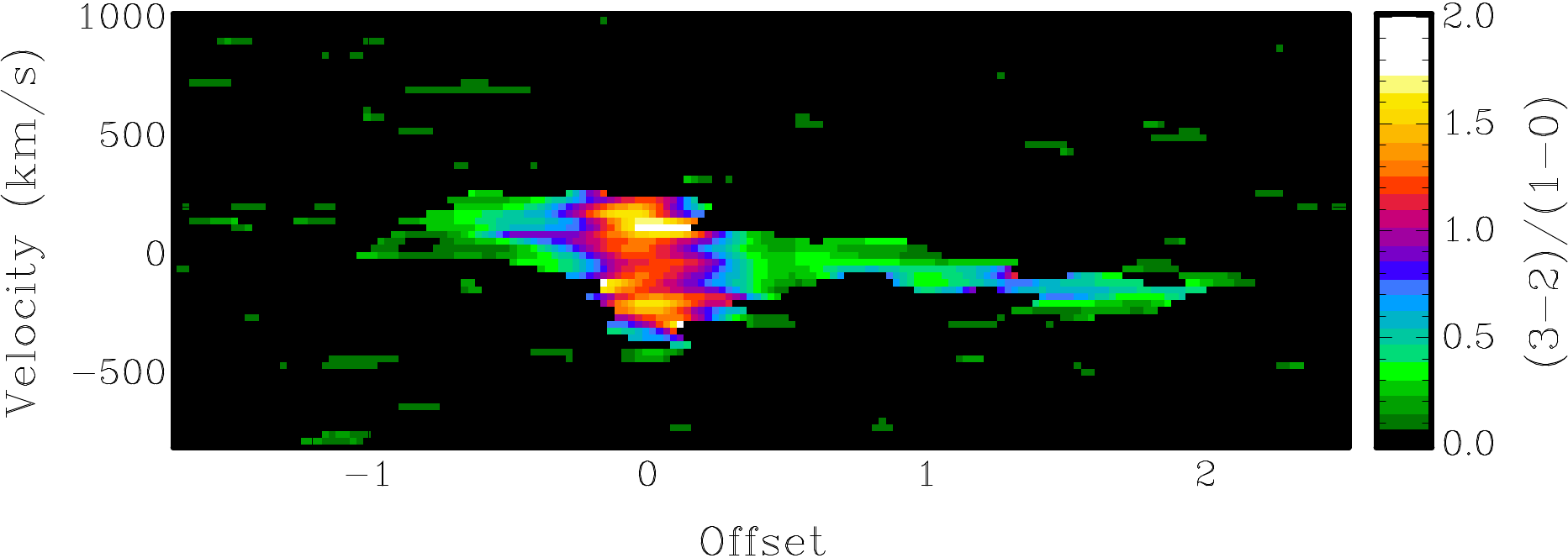}       
   \caption{Average N-S position-velocity slices of the \coOne\  (top) and \coThree\ (middle) which were used to compute the line ratios, shown at the bottom.  Contour levels are 0.4, 0.8, 0.12,... K (top) and 0.1, 0.4, 0.8, 0.12... K (middle). The direction of the slice is the same as in Figs \ref{fig:kpv} and \ref{fig:kpv32}.}
              \label{fig:ratSlices}%
    \end{figure*}

Following this, to improve the signal-to-noise of the data,  N-S position-velocity slices centred on the core were computed  from both data cubes by  spatially  averaging the data in the E-W direction   over a length of 0\farcs375 (1 kpc). These slices are shown in Fig.\ \ref{fig:ratSlices}. This figure underlines that the contrast between the inner disc and the outer tails is very different in the two transitions. One can  also see that the $pv$ diagram of the circumnuclear disc is somewhat different in the two transitions, with the very inner region with the higher velocities being much brighter in \coThree. This suggests that there is a gradient over  the circumnuclear disc in the excitation of the gas. 

This can be further seen in the bottom panel of Fig.\ \ref{fig:ratSlices} which shows the line ratio \rde\ = \coThree/\coOne\ (where the brightness used is in Kelvin) as computed from the $pv$-slices shown in Fig.\ \ref{fig:ratSlices}. The line ratio was only computed for those pixels where $S_{\rm CO(1-0)} > 0.4$ K.  In the outer tails,  \rde\ is well below 0.5, which is typical for the ISM in large-scale gas discs in galaxies (\citealt{Leroy09,Oosterloo17}). In contrast, in the inner regions \rde\ is  much larger. The average value in the core region of  \rde\ is 1.25, with a maximum for \rde\ of 2.3.

\begin{figure}
   \centering
   \includegraphics[width=6.5cm,angle=0]{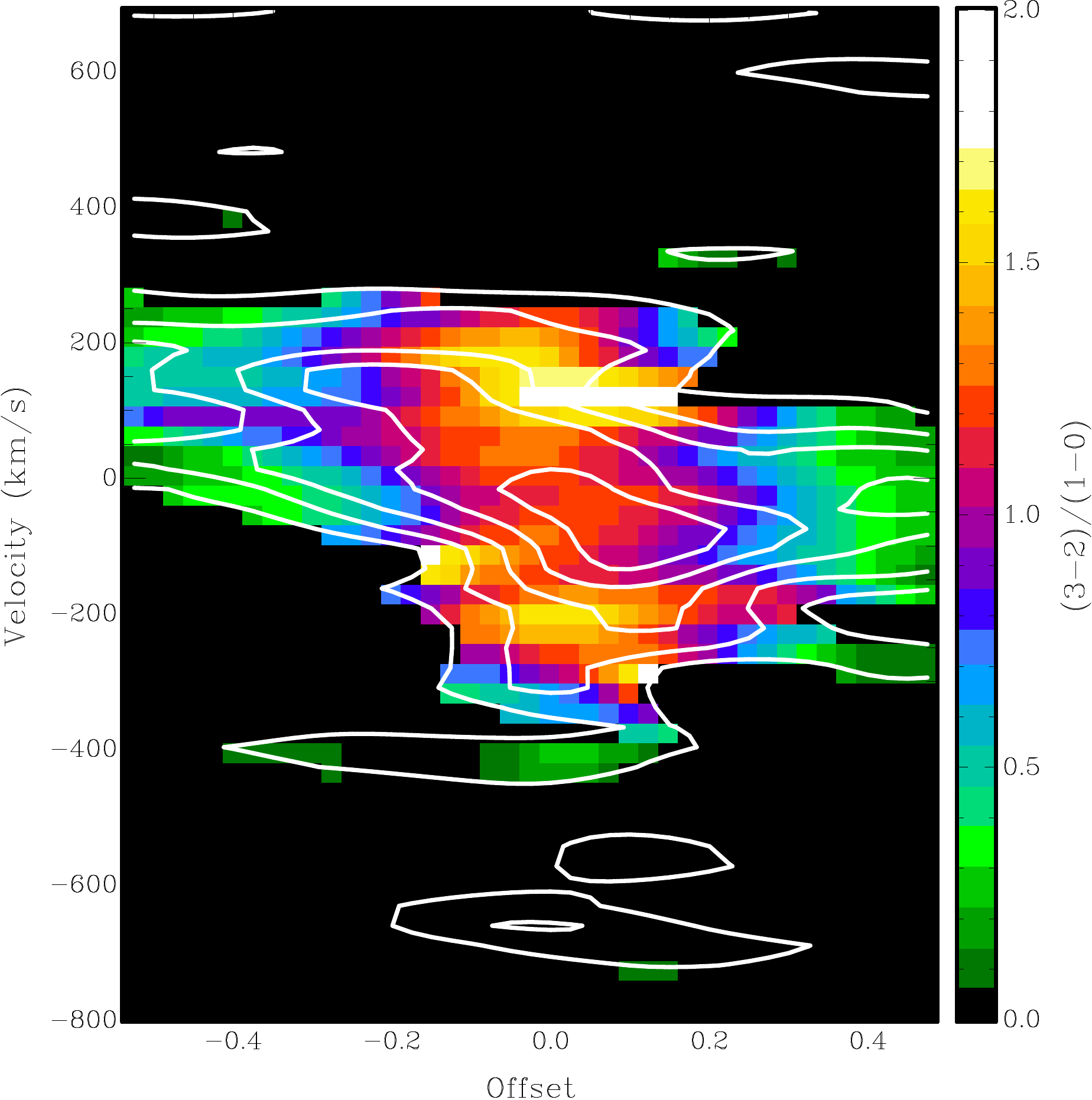}
   \caption{Zoom-in of the position-velocity shown in Fig. \ref{fig:ratSlices} bottom, illustrating the line ratio in the central region of \pks, clearly showing that the highest ratios are at the location of the core. Contour levels are 0.4, 0.8, 0.12,...K.}
              \label{fig:rats}%
    \end{figure}

The zoom-in in Figure \ref{fig:rats} shows that high values for \rde\ are mainly found in the inner parts of the circumnuclear disc, very close to the centre.  In this figure, the \coOne\ intensity contours of the circumnuclear disc have been overplotted. This  shows that the shape of the circumnuclear disc in \coOne\ in the $pv$-plane is different from that of the region of elevated line ratios. The intensity contours show the typical S shape of a $pv$ diagram of a rotating disc, while the region with high ratios shows this to a much lesser extent, implying that the gas with the highest line ratios must be in the inner parts of the disc. Similarly, Fig.\ \ref{fig:rats} also shows that the highest ratios occur at velocities away from the systemic velocity, meaning that the fastest moving gas has the highest excitation.

The high values for \rde\ ($> 1$) observed in \pks\ suggest that the CO emission from the circumnuclear disc is  optically thin. This is relevant for estimating the mass of the molecular gas in the inner regions (see below). In addition, they also indicate that the physical conditions of the molecular gas are characterised by much higher kinetic temperatures  and  densities than those found in a normal interstellar medium. This is likely caused by large amounts of energy being pumped into the ISM by star formations or by the AGN. Ultra Luminous Infrared Galaxies typically have \rde\ $\leq1$ \citep{Greve14}, although this is based on integrated fluxes, so that locally higher values may occur. However, values for \rde\ well above 1 are observed in some AGN. An example is NGC 1068 \citep{Viti14} where very elevated values for \rde\ are seen in the circumnuclear disc, where the gas is likely excited by the AGN. The models of \citet{Viti14} suggest densities up to $10^5$ cm$^{-3}$ and kinetic temperatures up to 150 K. A similar case is IC~5063 (\citealt{Dasyra16,Oosterloo17}), where the  CO gas that is kinematically disturbed by the AGN has very high values for \rde. Modelling the various line ratios in IC 5063 suggested that the disturbed gas  has   densities and kinetic temperatures similar to those observed in NGC 1068. Also in the case of IC 5063 this is very likely the result of the AGN dumping large amounts of energy in the ISM surrounding the AGN.


\subsection{The molecular gas masses}
\label{sec:molecularmasses}

The masses of the different components in the molecular gas  were estimated from the CO(1-0) cube. We estimate the mass of the three structures separately: the tail, the central disc and the broad component seen in CO(1-0).

The flux integral of the broad component was derived by integrating the spectrum at the position of the core over the velocity range $-350$ to $-1800$ \kms, resulting in a flux integral of 0.24 Jy \kms. For the central disc, we integrated the  signal in the data cube over a region of 0\farcs2 $\times$ 0\farcs2 and the velocity interval $-350$ to +350 \kms, obtaining a flux integral of 1.0 Jy \kms\ for this component. For the fainter, extended tails we obtain a flux integral of 1.85 Jy \kms.

It is likely that the emission of the central disc and of the broad component overlap in the data cube over a certain velocity range and it is difficult to separate the two. A rough correction would be to assume that  in the velocity range of the central disc, the broad component contains as much emission as it does in the blueshifted velocity interval not overlapping with the central disc. Making this assumption, the corrected flux integral for the central disc is 0.76 Jy \kms\ and for the broad component 0.48 Jy \kms. 

To convert these measurements into masses, one has to make assumptions about the conversion factor. For the extended tails we used a  standard conversion of 4.6 K \kms\ pc$^2$ because there are no indications that this gas has unusual excitation. This results in an  estimate for the H$_2$ mass of the tails of $9.7\times 10^9$ \msun. Given that the observed line ratios are larger than 1, the gas in the central regions is likely optically thin and has different excitation so a  much lower conversion factor may have  to be used. For a  conversion factor of 0.3 K \kms pc$^2$ representative of such conditions \citep{Bolatto13}, the mass in the broad component is $1.6 \times 10^8$ \msun\ and for the central disc  $2.6 \times 10^8$ \msun. It is interesting to note that the total H$_2$ mass derived from our observations is  close to  the $6.7 \times 10^9$ \msun\ estimated using the far-IR luminosity ($L_{\rm IR} = 1.6 \times 10^{12}$ $L_{\odot}$, \citealt{Holt06}) and the conversion between FIR and CO luminosity from \citet{Ocana10}.

These masses imply that the beam-average column density in the central region is  about $9\times 10^{22}$ cm$^{-2}$.
Interestingly, this is consistent with the neutral column density $N_{\rm H} = 5.2 \pm 0.1 \times 10^{22}$ cm$^{-2}$ found by \citet{Tombesi14} from X-ray observations.
In addition, by modelling the effect of extinction on the optical SED of \pks,  \citet{Holt06} derived an  \HI\ column density in the range $  1.2 \times 10^{22} < N_{\rm HI} < 2.4 \times 10^{22}$ cm$^{-2}$.  Thus, although each of the column density tracers we use here has its uncertainties and they are not necessarily expected to give the same answer (e.g.\ due to differences in geometry), the column densities derived from them are comparable. From the 21-cm \HI\ absorption, \citet{Holt06} derive  a column density of $N_{\rm HI} = 4.0\pm 0.5 \times 10^{18} T_{\rm spin}$ cm$^{-2}$. The  \HI\ column densities derived from the optical data and from the HI absorption are consistent if the  spin temperature of the absorbing gas would be in the range 3000--6000 K. This is not untypical for atomic gas in the immediate vicinity of an AGN due to the effects of the radiation field of the AGN on the excitation of the 21-cm line \citep{Morganti18}.  On the other hand, the small width of the \HI\ profile suggests that the gas is not near the AGN but more likely associated with the large-scale gas at larger radii.

\subsection{The molecular outflow}
\label{sec:results_energetics}

We use the time-average thin-shell approach $\dot{M} = M_{\rm out}V_{\rm out}/R_{\rm out}$ (\citealt{Rupke05}) to estimate the mass outflow rate of the molecular gas. It is not entirely clear what to use for $V_{\rm out}$ because the total width of the broad profile is the sum of the bulk outflow velocity and the turbulence of the outflowing material and it is unknown what the relative contributions are. For the purpose of this paper, we assume an outflow velocity of 600 \kms. Using these numbers, we find the outflow rate to be about 650 \msunyr. This number is, however, quite uncertain. In the first place because it is difficult to estimate the flux of the molecular gas involved in the outflow due to the overlap of the  emission form  the outflowing gas and from the circumnuclear disc  in the 3-D data cube. Secondly, the conversion factor to use to convert fluxes to masses is quite uncertain and the value used here likely represents a lower limit. In addition, the outflow velocity may well be larger than our conservative estimate, while we can also only set an upper limit to the projected size of the outflow region.  Projection effects could play a role and the true size of the outflow region, if the outflow is quite collimated and directed close to the line of sight, might be larger.  Despite all these uncertainties, it is clear that the molecular outflow is  much more massive than the one seen in the ionised gas which has an outflow rate of $\dot{M} < 10$ \msun\ (\citealt{Holt06}; Santoro et al.\ in prep.). 

The kinetic power  associated with the molecular outflow can be estimated using
$\dot E = 6.34 \times 10^{35} (\dot{M}/{2}) (v^2+3\sigma^2)$ (\citealt{Rodriguez13,Mahony16}) where $v$ is the velocity of the large-scale motion and $\sigma$ the turbulent velocity. Our observations do not allow us to pin down accurate values for $v$ and $\sigma$ separately, however the kinetic power depends on the combination of the two and assuming a lower value for  $\sigma$ would have to be offset by a larger assumed value for $v$.
To obtain a rough estimate, we assume $v = 600 $\kms\ and $\sigma = 510$ \kms\ (FWHM = 1200 \kms), giving an estimated kinetic power of the molecular outflow  of a  few times $ 10^{44}$ \ergs. 
\citet{Holt06} estimated the bolometric luminosity of \pks\   to be between $9\times 10^{45}$ \ergs\ and  $4 \times 10^{47}$ \ergs, while Santoro et al.\ (in prep.) estimate the bolometric luminosity to be $6\times 10^{45}$ \ergs.
Thus, the ratio between the kinetic energy carried away in the molecular outflow and the bolometric luminosity is of the order of a few per cent.

\section{Results: properties of the continuum emission}
\label{sec:results_continuum}

\subsection{The structure of the radio continuum}

\begin{figure}
   \centering{
       \includegraphics[width=8cm,angle=0]{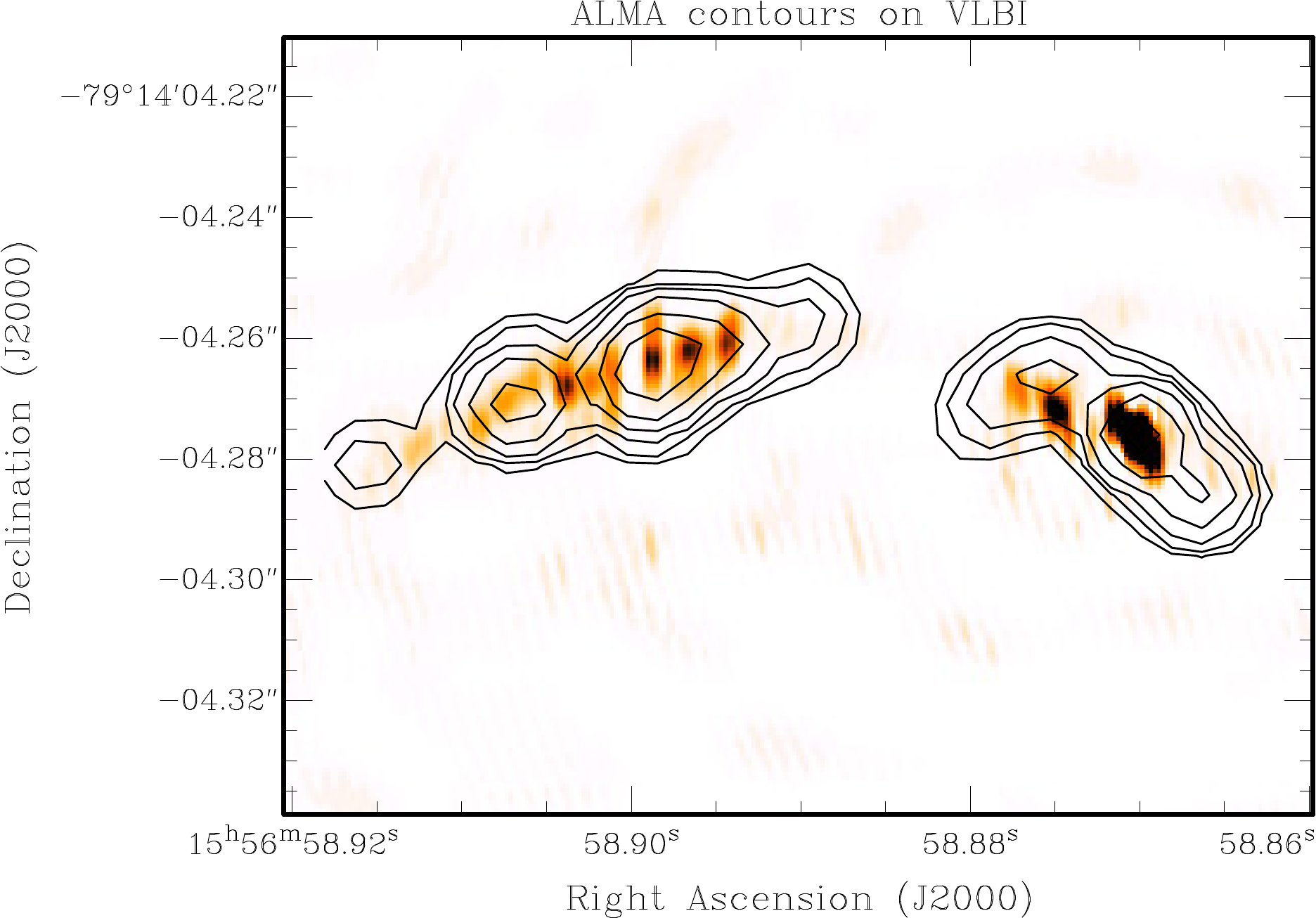}}
   \caption{VLBI 2.3-GHz continuum (grey scale) superposed on  the super-resolved 100-GHz image (black contours). Contour levels are 0.15, 0.3, 0.6, 1.2 and 2.4 \mJybeam.}
              \label{fig:vlbicontinuum2}
    \end{figure}

Figure \ref{fig:vlbicontinuum} shows our 2.3-GHz radio continuum image as obtained from our VLBI observations while Fig.\  \ref{fig:vlbicontinuum2} shows the comparison of this image with our super-resolved 100-GHz image.  The structure recovered in the new VLBI image is, to first order, similar to that in the VLBI image presented by \citet{Holt06} based on observations performed between November 1988  and March 1992. It should be noted that this latter VLBI image was obtained using a smaller  bandwidth (2 MHz) and single polarisation, therefore having much lower sensitivity. 

It is clear that the structure of the continuum source is strikingly similar at 2.3- and 100 GHz, despite the factor 50 difference in frequency. This strong similarity shows that  at mm wavelengths the radio continuum is dominated by non-thermal emission.  
The radio continuum emission of \pks\ consists of a strong central nuclear region with a strong core and a small jet with a position angle of about 45$^\circ$. A small counterjet in the opposite direction is also detected near the core.   In addition,  a large jet-like structure  is present, eastwards of the core,  extending to about 120 mas (300 pc) and having a different orientation than the inner jet, giving the overall appearance of a bent radio structure. The two features do not appear to be connected, with a gap of about 30 mas (80 pc) in between them. 

Interestingly, our new VLBI image shows that the nuclear region is relatively symmetric, showing a jet and a counter-jet. The counter-jet is also visible in the high-resolution 8.4~GHz image of \cite{Ojha10} and    in the super-resolution ALMA 100~GHz image. On the other hand, on larger scales, no emission is seen on the western side of the core and the radio structure is very asymmetric. This could be caused by the  jet being more or less aligned with the line-of-sight so that relativistic beaming effects play a role as suggested in earlier studies  (e.g.\ \citealt{Holt06}), but it could also be caused by an asymmetric interaction between the jets and the ISM. 
To shed some light on this,  we compared our VLBI image with the 2.3~GHz VLBI image   presented in \citet{Holt06}  (see also \citealt{Tzioumis02} for more  technical details) made from data taken between November 1988 and March 1992, about 17 years before our observations. Any change in the structure due to super-luminal motion over these 17 years   would indicate that the jet is quite aligned with the line of sight.

Although some details appear different in the two images, no shift is seen in the position of the knots along the jet at the level of about 5 mas (i.e.\ 3 beam sizes in the E-W direction). Because of this, we conclude that, at least for the large-scale jet, we do not see evidence of superluminal motion to explain the strong asymmetry observed on the hundred-pc scale. In this respect, the detection of a counter-jet in the nuclear region  is interesting because this  makes the hypothesis of a the jet being along the line-of-sight less compelling than previously thought. This, together with the bending of the large scale  jet,  suggests that instead the morphology of the radio continuum is affected by a strong interaction of the jet with the surrounding rich ISM.

\subsection{Spectral index}
\label{SpectralIndex}

The integrated spectral index ($\alpha$, here defined by $S \sim \nu^{\alpha}$) of \pks\ between 2.7 and 4.8 GHz was found to be relatively flat $\alpha = -0.17$ by \cite{Morganti93}. 
However, given that the observations were taken at different epochs, the effect of  radio variability cannot be excluded  (as also noted  by \citealt{Holt06}). Indeed, looking at the data taken at different times and frequencies  as part of the monitoring campaign of ATCA calibrators\footnote{See the ATCA calibrator database, \url{https://www.narrabri.atnf.csiro.au/calibrators/calibrator_database_viewcal?source=1549-790}} (of which \pks\ is part), one can see that the spectral index between 2.1 and 5 GHz has become increasingly steeper in  recent years. The  flux density of \pks\ at 5 GHz has decreased by about a factor 2 between the observations done in 1991--1992 as presented in \citet{Morganti93} and monitoring measurements done in 2016. This suggests that the activity of the core, the component providing the dominant contribution to the flux, is changing quite dramatically on relatively short time scales and, in addition, that other structures, such as the  radio jets, provide a relatively larger contribution to the spectral index at later times.
The change in the core activity could be taken as a signature of the intermittent fuelling of the SMBH, possibly as result of feedback.

Furthermore, a clear curvature is present in the integrated spectrum, with the spectral index steepening at higher frequencies and a flattening observed at low frequency (below a few hundred MHz) when data from the GLEAM survey at 150 MHz \citep{Hurley17} are considered. Curvature of the spectrum at such low frequencies, if due to synchrotron self-absorption, is usually associated with older sources with characteristic sizes of a few kpc (\citealt{Snellen00}),  much larger than \pks. This could be an indication that the source is actually older than what the size would suggest, possibly due of the confining action of the rich gaseous medium.

The steepening at  high frequencies is further confirmed by the fluxes derived from the ALMA data. Furthermore, using the VLBI- and the super-resolved ALMA data, we can separate the emission of the core region (including the inner jet and the counter jet) from that of  the more extended jet and derive their respective spectral indices  (see Table \ref{tab:fluxes}).

Table \ref{tab:fluxes} shows that there is a clear difference between the spectral index of the nuclear region and that of the jet: between 2.3 and 100 GHz, the spectrum of the extended jet a few hundred pc from the core is steep ($\alpha \sim -1.15$)  while the core region has  a  flatter spectrum ($\alpha \sim -0.6$). Steeper spectra are seen between 100 and 300 GHz for both components.  A  difference between the spectral indices of the core and of the large-scale jet was already noted by \cite{Holt06}, although we find a steeper spectral index for the core region than \citet{Holt06}. We reiterate that the spectral index of the core region should be taken with care because it could be affected by variability.

The steep spectral index of the jet found by \citet{Holt06} is confirmed by  our data. A very steep spectrum  (steeper than about --1.2) is often associated with dying structures where the energy injection by the active nucleus has stopped. However,  for a limited number of sources that are still relatively young (and small), very steep spectrum structures  have been detected. 
In particular, fader sources (e.g.\ 1542+323, \citealt{Kunert05}; 0809+404, \citealt{Kunert06}) are possible examples of young radio sources that are dying. A well studied case is  PKS 1518+047 (\citealt{Orienti10}), where the entire structure is characterised by a very steep spectrum and the core is lacking. An other possible example is PKS B0008--421 (\citealt{Callingham17}).

In the case of \pks, we may be seeing a dying remnant structure (the jet on hundred-pc scale) not being fed by the AGN anymore.
This could be the result of a temporary disruption of the jet, or of intermittent fuelling of the AGN, both due to a strong interaction of the jet with the rich medium in which the AGN is embedded. 

\begin{table} 
\caption{Continuum flux densities and spectral indices derived from the ALMA and VLBI  observations. }
\begin{center}
\begin{tabular}{cccccccc} 
\hline\hline 
   & $S_{\rm 2.3\ GHz}$ & $S_{\rm 100\ GHz}$  & $S_{\rm 300\ GHz}$ & $\alpha^{2.3}_{100}$ &$\alpha^{100}_{300}$\\
     &  (Jy) & (Jy)  & (Jy)\\
\hline
 Core   & 3.02  & 0.354 & 0.147& --0.57 & --0.80\\
 Jet  & 1.78 &0.023 & 0.0038 & --1.15 & --1.64\\
\hline
\end{tabular}
\end{center}
\label{tab:fluxes}
\end{table}

\section{Accreting and outflowing molecular gas}
\label{sec:discussion}

\subsection{Overall properties of the molecular gas}

The high-resolution observations of the molecular gas that we  present in this paper help to derive a picture of the crucial early phases of the evolution of \pks, and of the role the AGN plays in this. 
Earlier observations had revealed several interesting features in \pks\ which we briefly summarise here, following the results presented in \citet{Holt06} and references therein. Optical observations had  shown the presence of large-scale tail-like structures in the ionised gas and in the stellar distribution, indicating  a recent merger has occurred and that this merger is still in progress. The high FIR luminosity of \pks\ indicates that this merger is gas rich and a that large amount of star formation is associated with this merger.  In addition, the data showed that \pks\ contains a highly reddened AGN, associated with a small radio source, which must be obscured by a large amount of gas in the central regions.  A newly born radio jet  was detected which is fighting its way out the dense, rich medium of the merger remnant and which appears to  drive an outflow of warm gas, as detected through blue-shifted, very broad optical emission lines, although the energy carried by this warm outflow is relatively small and  is not capable of clearing the central regions. 

This picture is further confirmed and expanded by our CO  observations. 
Overall, our data show that the merger drives large amounts of molecular gas towards the central regions where this gas feeds strong star formation and where some of the gas is settling in a circumnuclear disc. Part of this central  gas reservoir is  able to feed the active super-massive black hole. On the other hand, a strong outflow of gas occurs in the very centre. Given that this outflow occurs only in the very inner regions suggests that the AGN drives the outflow. Therefore feedback and fuelling co-exist and interact with each other. The variability and other properties of the radio continuum suggest that the fuelling of the AGN is intermittent, possibly as the result of a continuously changing balance between feedback and fuelling.

On the larger, kpc scales, we detect a large amount ($\sim10^{10}$ \msun) of molecular gas. This  is consistent with the ULIRG/FIR properties of \pks\ and its large amount of star formation. Our observations also show that, on these scales, the  molecular gas  forms two tails, which mirror those seen for the stars and for the ionised gas. The total extent of the northern tail is about 1\farcs5 arcsec (about 4 kpc), while the length of the southern tail is about 0\farcs5 arcsec (about 1 kpc). Interestingly, although the linear structure these tails form on the sky suggests that  we see these tails fairly edge on, the velocity gradient we detect over them is very small, suggesting the kinematics on the largest scales is dominated by radial motions, in the plane of the sky. This could indicate a large-scale inward gas flow  which would lead to  gas piling up in the central regions. 

In the inner regions, the brightness distribution and the kinematics of the CO gas is very different and the gas forms a distinct component there, which is rotating about the centre. In particular, the \coThree\ is  bright in the inner few hundred parsec and the velocity width of the emission there is much larger than in the tidal tails. The kinematics in the central regions appears more settled, being quite symmetric with respect to the centre  and with clear signs of rotation. All this suggests that  in the inner regions, the molecular gas forms a circumnuclear disc with a radius of about 240 pc and which is seen at fairly high inclination. In projection, this disc runs north-south over the nucleus, perpendicular to the jet axis and covering the AGN. The observed column densities of the circumnuclear disc are consistent with those derived from optical and X-ray observations, suggesting that the strong extinction seen in the optical spectrum of the AGN is coming from this circumnuclear disc.

The impact of the AGN on the  gas of the circumnuclear disc in \pks\ becomes clear from studying the line ratio \rde\ of the molecular gas. Similar to what is seen for other AGN, the gas in the direct vicinity of the AGN shows high values for the line ratio (\rde\ $> 1$), very different from the gas in the large-scale tails at larger radii which shows  ratios typical for a normal ISM. The high values observed  in the central regions imply that the molecular gas near the AGN is optically thin and has elevated excitation temperatures. Close inspection of the distribution of \rde\ in the data cube suggests that the inner parts of the circumnuclear disc have the highest values of \rde. In this inner region,  it is also the gas with the highest velocities with respect to the systemic velocity which has the highest line ratios, however this could be partly due to overlap in the data cube of  AGN-affected emission from the circumnuclear gas and the quiescent large-scale tails.

Although the effect of the AGN on the cold ISM in \pks\ appears to be limited to the inner regions, the impact there is likely very significant and with the observed mass outflow rates, the circumnuclear disc could be destroyed on a relatively short time scale. The gas mass of the circumnuclear disc is a few times $10^8$ \msun, while the mass outflow rate is at least 650 \msunyr. This means that on a time scale of  $\sim$10$^5$ yr the AGN would be able to destroy the central disc. On the other hand, gas from large radius is flowing towards the centre, providing material for rebuilding the disc.

\subsection{Possible scenarios for the outflow} 
\label{sec:energetics}

The molecular outflow we detect in the form of a very broad CO profile at the position of the AGN is likely the molecular counterpart of  the ionised outflow detected earlier by \citet{Holt06}. 
From our data we cannot unambiguously derive which mechanism (starburst, wind, radiation pressure or radio jet) is causing these phenomena. Indeed, all candidate processes are present in \pks: a strong starburst is occurring in \pks\  and furthermore, of the objects studied so far where molecular outflows have been detected, \pks\ is one of the strongest radio sources. At the same time, it harbours a powerful optical AGN \citep{Holt06}. The existence of an Ultra Fast Outflow (UFO) with a speed of about $0.28 c$ is detected in the X-ray spectrum  \citep{Tombesi14}, which suggests that a wind from the accretion disc is also present.

   \begin{figure}
   \centering{
       \includegraphics[width=8cm,angle=0]{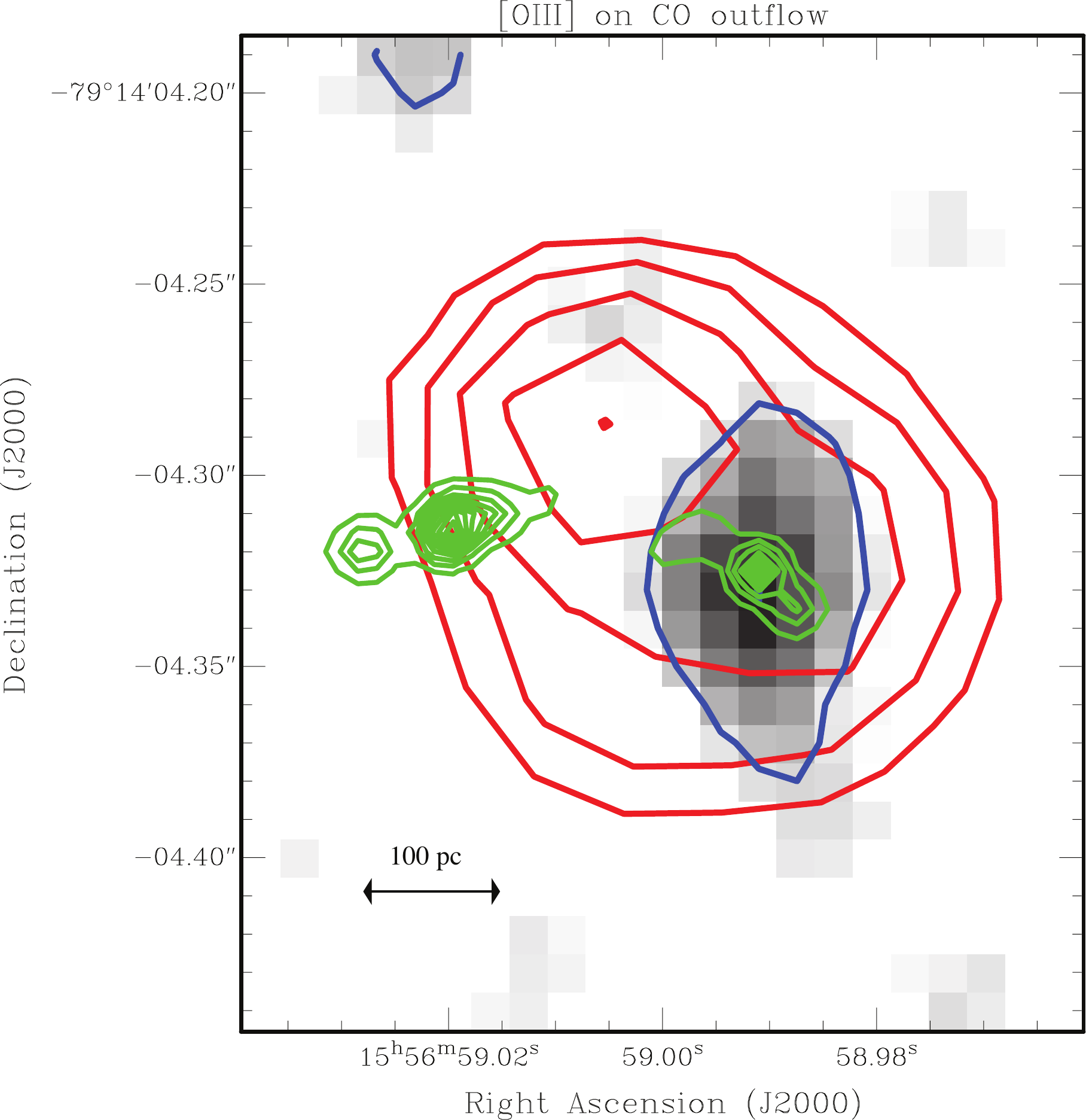}}
   \caption{Relative extent of the \OIII\ emission (HST data from \citealt{Batcheldor07}; red contours) compared to the super-resolved 100-GHz radio continuum (green contours) and the region of the broad \coOne\ profile (blue contours).  The \coOne\ distribution was obtained by integrating the \coOne\ data cube over the velocity range --300 \kms\ to --2000 \kms. Contour levels for the HST data are 0.25, 35.4, 50, 70.7 and 100\% of the peak emission while for the \coOne\ they are 50 and 100\% of the peak.}
                  \label{fig:HSTcomp}

    \end{figure}
In Fig.\ \ref{fig:HSTcomp} we compare the relative extent of the \OIII\ outflow seen in the HST image of \citet{Batcheldor07}, the region of the CO outflow and the radio jet\footnote{The optical spectroscopy of \citet{Tadhunter01,Holt06} and Santoro et al.\ (in prep.) shows that the \OIII\ emission integrated over a large aperture centred on the AGN is dominated by the outflow component, with the entire \OIII\ emission-line profile shifted by several 100 km s$^{-1}$. Therefore we can be confident that the HST narrow-band \OIII\ image maps the warm outflow}. We note that, based on the accurately measured offset
between the quasar nucleus that dominates the K-band image of \citet{Inskip10} and a nearby star that is detected in both the K-band and the optical HST images, the AGN is not
centred on the brightest part of the \OIII\ emission, but rather on a secondary peak $\sim$0\farcs07  (190 pc) to the southwest. Thus, the brightest \OIII\ emission could represent the site of a past jet-cloud interaction that has deflected the jet. However, the jet is unlikely to be currently interacting with the cloud, else we would expect to detect a bright radio knot close to the peak of the \OIII\ emission.
Based on these HST data, \citet{Batcheldor07} concluded that there is no evidence for  bi-conical  emission-line features which one would expect for an outflow driven  by strong star formation.  Figure \ref{fig:HSTcomp} indeed shows that the \OIII\ outflow is one-sided.  However,  star-formation driven outflows can be  asymmetric. In addition, extinction effects, which are known to play a role in \pks\ \citep{Holt06}, could make the ionised outflow appear much more asymmetric than it actually is. Hence, the one-sidedness observed for the ionised outflow does not completely exclude it is driven by a star burst.  The warm outflow emerges from the nucleus with the same position angle as the inner radio jet which suggests a link between the jet and the outflow, but the radio-optical agreement deteriorates as the jet curves around while the orientation of the \OIII\ outflow does not seem to change there.

Interestingly, Fig.\ \ref{fig:HSTcomp} also shows that the spatial extent of the \OIII\ outflow is larger than that of the CO outflow. The \OIII\ outflow extends to about a radius of 0\farcs07 whereas the CO outflow is unresolved in our data ($r < $ 0\farcs045).  The fact that the molecular outflow only occurs within $r < 120$ pc may argue for the AGN to be driving the outflow, because the strong star formation is likely to happen over a larger region.

The stratification  we may be seeing between the molecular- and ionised outflow could fit with a scenario where a jet that is working its way through a clumpy ISM is responsible for the turbulence and the outflow, as has been modelled recently using simulations of a jet moving through a clumpy ISM \citep[e.g.][]{Mukherjee18a}. 
In the initial stages of such an interaction, the progress of the jet  is intermittently blocked by the denser clumps in the ISM, causing the jet to meander through the ISM, moving from dense cloud to dense cloud. This is suggested  by the results on the morphology and spectral index of the radio continuum of \pks\ described in Sec.\  \ref{sec:results_continuum}. The interaction process can be highly asymmetrical as it depends on the detailed local clumpy structure of the ISM. While the jet is temporarily blocked, large amounts of energy are dumped by it in the ISM.  According to the simulations, in this way, the jet produces a cocoon of shocked gas of mixed density and temperatures, as well as a backflow which expands in all directions through the ISM. This   induces strong turbulence in the ISM in all directions, including in the molecular gas. After the jet has meandered through the ISM for a while, the over-pressured cocoon of warm/hot gas  breaks out from the clumpy ISM and a strong outflow of warm/hot gas is created into the halo of the galaxy, perpendicular to the ISM disc. The orientation of this second phase is more or less independent of the relative orientation of the jet and disc. 
In the models, the time scale of the effects depends on the relative orientation of the jet and the disc, being longer if the jet is oriented in the plane of the disc, but even if the jet is perpendicular to the denser disc, strong interactions and outflows occur in the ISM. To some extent, in this kind of model,   two kinds of feedback occur. One is the direct interaction of the jet with dense gas clumps, directly affecting the denser ISM near the AGN. The other  is dumping energy, through the warm outflow, in the larger-scale, less dense gaseous halo of the galaxy, increasing the time scale over which the gas in the halo can cool and form stars (maintenance mode feedback).

At least qualitatively, this model may describe the outflows we see in the molecular- and in the ionised gas. The molecular outflow would come from the region of direct feedback and would correspond to that part of the interaction where the jet-inflated cocoon is driving turbulence in the denser gas disc. The relatively small extent of  the region showing elevated line ratios in the molecular gas suggests that  the denser ISM  is affected  out to about 0.5 kpc radius.     The turbulent/outflowing molecular gas could represent denser clumps in the jet-driven outflow that have short cooling times, or clouds that have had more time to cool following an earlier interaction with the turbulent jet cocoon (earlier since closer to the nucleus). 

In contrast, the more extended emission-line outflow detected in \OIII\  could represent gas that has broken out of the disc and is currently flowing into the lower density halo of the galaxy. This gas  has perhaps been accelerated and ionised by a shock induced by the inflating jet cocoon. The fact that this gas is still ionised and has not yet cooled to a molecular phase could be because it has a lower density and hence a longer cooling time. Alternatively, it has a higher density, but has been accelerated in a more recent interaction with the expanding jet cocoon and has not yet have time to cool. The latter would be consistent with the high densities observed in the \OIII\ outflow in \pks\ from the X-shooter spectrum presented in Santoro et al.\ (in prep). 
The molecular gas in the large-scale tails at much larger radii (several kpc) appears to be unaffected by the AGN (so far) and it is unlikely that the evolution of the gas on these larger scales, and the star formation from this gas, will be changed by the AGN activity.

If what we see in \pks\ represents a  phase in galaxy evolution common to many galaxies, it is useful to compare the results presented here with what found for other objects available in literature. 
A number of cases are known now for which observations suggest  the scenario of a jet interacting with dense clumps in the ISM is  happening in many objects. Interestingly, these include AGN of both low- and high radio power.
In \citet{Husemann19a}, a rich set of multi-wavelength observations are presented of the low-power AGN HE~$1353-1917$. 
In this object, the jet is moving in the plane of the disc and is driving a multi-phase outflow over a region of about 1 kpc. Using observations in several wavebands, \citet{Husemann19a} were able to show that  it is most likely that the plasma jet is responsible for this. Another recent case is the well-known quasar 3C~273 \citep{Husemann19b} where the data suggests the presence of an expanding over-pressured  cocoon of hot gas created by the powerful radio jet which impacts on an inclined gas disc and which drives fast transverse and/or backflow motions.  
 
Another  case for which detailed information is available is IC 5063 \citep{Morganti15,Oosterloo17}, where the close morphological match between the radio jet and  the region of molecular gas with elevated line ratios  and extreme kinematics  is a strong indication that the interaction between the radio plasma with the ISM is the main mechanism for disturbing the gas and for affecting its excitation. In HE 1353--1917 and IC 5063, the highest excitation and the most extreme kinematics do not occur at the core, but along the entire jet where it appears to be interacting strongly with the ISM. The spatial resolution of our \coThree\ data on \pks\ is not sufficient to investigate in great detail whether similar processes occur in \pks, but there are hints that in \pks\ the situation is different and that instead the AGN-affected gas is found near the core and not along the jet. There are no features in the CO morphology and kinematics along the jet  that would suggest that the conditions  are peculiar there. Instead, both the \coOne\ and \coThree\ clearly peak at the core and   the highest line ratios, as well as the kinematically disturbed gas, are observed there. One reason for these differences could be the large disparity in jet power between HE 1353--1917  and IC 5063 on the one hand and \pks\ on the other. The simulations of \citet{Mukherjee18a,Mukherjee18b} show that a low-power jet may affect a very large region of the dense ISM because the jet takes much more time to break through the ISM, while a more powerful jet pierces through the dense ISM relatively more quickly. In addition, the fact that in HE 1353--1917  and IC 5063 the orientation of the jet is such that the jet moves in  the plane of the gas disc, while in \pks\ the jet is not aligned with the disc, may play a role in explaining the differences. Perhaps due to this, in \pks\  the outflow, despite the more powerful AGN, is limited to the inner $\sim$120 pc. 
Common among all examples mentioned here is that the radio jet seems to have an impact up to a few kpc from the AGN and less so on its overall ISM (see also \citealt{Murthy19}).

Despite the relatively small region affected in \pks, the molecular outflow rate is large, at least 650 \msunyr, and much larger than that seen for the ionised gas. This is similar to other cases of obscured quasars, or to galaxies observed to be in the process of quenching and where molecular outflows have been found (e.g.\ \citealt{Sun14,Brusa18,Herrera19,Veilleux17}).   \citet{Fiore17}, albeit based on a heterogeneously selected sample, have claimed that the difference between ionised- and molecular outflow rates decreases for the most luminous AGN ($L_{\rm bol} > 10^{46}$ \ergs).  An example is the powerful obscured quasar XID2028 for which \citet{Brusa18} found similar mass outflow rates for the ionised and the molecular gas. 

However, the reported trends do not take into account that in some objects the radio jets can be the dominant driver of the outflow, instead of the AGN luminosity, so the situation may be more complicated. In the case of \pks, despite being a powerful optical and radio AGN, we find a large difference in mass outflow rate between the molecular and the ionised outflow. This suggests that either the trends with AGN luminosity are more complicated, or a more prominent role for different mechanisms, such as the radio jet (for which the connection between jet power and outflow parameters  is still not properly investigated).  

Statistical studies of representative samples are still scarce. The only survey of molecular gas in similar objects (local ULIRGs and QSO’s) is the one of \citet{Cicone14} where they confirm the high incidence of molecular outflows.  The situation is less clear for dust-obscured galaxies (DOGs) with only a few (high redshift) objects studied so far with ALMA  and where contradictory results were obtained, although not many outflows were detected (see \citealt{Toba17,Fan18}). 
A common conclusion is that in all cases studied in detail,  the gas depletion times of the inner region are relatively short ($\sim 10^5$ - $10^6$ yr)  and, therefore, these massive outflows are representing a relatively short (but likely recurrent) phase in the evolution of these objects.  These short depletion times may be connected to observations of AGN that seem to be dying or flickering on very short time scales (e.g.\ \citealt{Ichikawa19,Schawinski15}).

\section{Conclusions}

With our ALMA \coOne\ and \coThree\ observations of the obscured young radio quasar \pks, we detect the presence of a large amount of molecular gas ($\sim$10$^{10}$ \msun) in this object. 
The data show that the distribution and the kinematics of the gas is complex. We detect three distinc components:  extended gas tails related to an ongoing merger, a circumnuclear disc,  and a fast molecular outflow.

The large-scale tails of molecular gas reach into the central regions, feeding the large starburst occurring there. In the inner few hundred pc, the large, regular velocity gradient suggests the presence of a disc-like structure with a molecular gas mass of a few $\times 10^8$ \msun. All this is likely connected with  feeding  the growth of the central SMBH, which is known to be accreting at a high Eddington ratio (\citealt{Holt06}). Interestingly, the observations show that a nuclear disc manages to form, despite the presence of a powerful AGN disturbing the  gas in the inner regions, suggesting that feeding and feedback can co-exist.

The effect of the AGN  is seen in the form of  a fast and massive outflow detected in CO(1-0). We estimate a mass outflow rate of at least 650 \msunyr, possibly substantially higher. This is much larger than the mass outflow rate detected for the ionised gas (\citealt{Holt06}; Santoro et al.\ in prep.). The impact of the AGN on the gas is also confirmed by the  higher excitation of the molecular gas in the very central region as derived from the line ration \rde. The molecular outflow  could destroy the circumnuclear disc on a time scale of only $10^5$ yr, although gas from the merger is moving in radially, possibly (partly) rebuilding the disc at the same time.

Despite the presence of a powerful radio source and a quasar nucleus, the  massive outflow is confined to the inner region (radius $< 120$ pc) of the galaxy. The region of the outflow of ionised gas present in \pks\ appears to be more extended  than that of the molecular gas.  No AGN related effects are seen at radii larger than 0.5 kpc and most of the ISM in \pks\ is unaffected by the AGN. 

The data are consistent with recent numerical models of a young plasma jet interacting with a clumpy ISM. In such models, the progress of the jet is  blocked intermittently by dense clumps, leading to large amounts of energy being dumped in the ISM. This results in an over-pressured cocoon of gas with a wide range of densities and temperatures, large amounts of turbulence in the ISM, and an outflow of gas.   The molecular gas affected by turbulence/outflow may correspond to the denser clumps in this cocoon so they have a shorter cooling time. At some point in time, the hot/warm gas of the cocoon breaks out from the dense ISM, into the less dense halo of the host galaxy. The ionised outflow may correspond to this phase. Alternatively (or additionally), the disturbed molecular gas may be gas closer to the AGN so that it was affected by the jet earlier and thus has had more time to cool while the ionised outflow is gas at larger radius so that it is affected by the jet later and thus has had less time to cool.
This warm outflow will dump energy gaseous halo on larger scales and may help prevent cooling of the halo gas  which may reduce future star formation \citep[e.g.][]{Costa18}.

In summary, the data show the complexity of feeding and feedback in action at the same time in  the inner regions. On the other hand, the AGN does not seem to have a large impact on the overall ISM on the largest scales and the feedback effects are limited to the central few hundred parsecs.

A comparison of the properties of a, still small, group of AGN for which good data on both the warm, ionised-  and the cold, molecular outflows are available, shows similarities with \pks, but also a number of differences. Like in IC~5063 and HE~$1353-1917$, the AGN appears to affect not only the kinematics, but also the physical conditions of the surrounding gas. 
This is despite the more than two orders of magnitude difference in  radio power between \pks\ and these  two objects. Improved  statistics on the effects of differences in radio power of the jet, as well as  their inclination with respect to the distribution of the gas, is needed to properly quantify the impact of radio jets.
Furthermore, the molecular outflow, despite being limited to a small region, appears to carry most of the outflowing gas. This does not agree with the findings of \cite{Fiore17} which suggest  that the difference between warm ionised and molecular outflow rates decreases for the most luminous AGN ($L_{\rm bol} > 10^{46}$ \ergs). Considering the high bolometric luminosity of \pks, the two order of magnitude difference between the warm ionised and the molecular outflows suggests that understanding the relation between AGN luminosity and mass outflow rate requires a better understanding of the driving mechanism (i.e.\  wind/radiation vs radio jet), something that has not been taken into account so far.

\pks\   represents a rare, local example of an obscured quasar, but at higher redshift, such objects are now being detected in increasing numbers on the basis of their extreme Spitzer/Wise mid-IR colours (e.g.\ Dust-Obscured Galaxies or Hot DOGs, \citealt{Wu12}), and at least some of them are radio-loud (\citealt{Lonsdale15}). 
Like \pks, these sources are still enshrouded in their natal cocoon of gas. Our results demonstrate the potential of ALMA molecular line observations for understanding the evolution of such objects.

\begin{acknowledgements}
We thank Katherine Inskip for providing the fully reduced VLT/ISAAC K-band image of \pks\ and the surrounding field that was used to
improve the astrometric registration of the optical HST and and mm ALMA images. We also thank Joe Callingham for providing the GLEAM data.
This paper makes use of the following ALMA data: ADS/JAO.ALMA\#2017.1.01571.S. ALMA is a partnership of ESO (representing its member states), NSF (USA) and NINS (Japan), together with NRC (Canada), MOST and ASIAA (Taiwan), and KASI (Republic of Korea), in cooperation with the Republic of Chile. The Joint ALMA Observatory is operated by ESO, AUI/NRAO and NAOJ.
The Australia Telescope Long Baseline Array is part of the Australia Telescope National Facility which is funded by the Australian Government for operation as a National Facility managed by CSIRO.

\end{acknowledgements}

%
%

\end{document}